\documentclass[% 
reprint,
 amsmath,amssymb,
 aps,
floatfix,
]{revtex4-1}
\usepackage{graphicx}% Include figure files
\usepackage{dcolumn}% Align table columns on decimal point
\usepackage{amsmath}
\usepackage{amssymb}
\usepackage{bm}% bold math
\usepackage[bbgreekl]{mathbbol}
\usepackage{color} % Colours
\usepackage{braket}
\usepackage{gensymb}
\usepackage[dvipsnames]{xcolor}
\usepackage{lpic}
\usepackage[percent]{overpic}
\usepackage{stackengine}
\usepackage{physics}
\usepackage{siunitx}
\usepackage{textcomp}
\usepackage{slashbox}
\usepackage{lipsum}
\usepackage{hyperref}
\hypersetup{
    colorlinks=true,
    linkcolor=blue,
    filecolor=magenta,
    citecolor=blue,
    urlcolor=black,
}
\usepackage[mathscr]{eucal}
\usepackage{ulem} % this enables to cross out words
\usepackage{amsfonts}
\usepackage[colorinlistoftodos]{todonotes}
\usepackage{mathtools}% Loads amsmath
\usepackage{empheq}
\usepackage{esvect}
\usepackage{rotating}

\usepackage{xr}
\DeclareSymbolFontAlphabet{\mathbbm}{bbold}
\DeclareSymbolFontAlphabet{\mathbb}{AMSb}

\begin{document}

\preprint{APS/123-QED}

% \title{Floquete description for a Voigt based 3D magnetometer}
%\title{Synchronous microwave spectroscopy on optically pumped magnetometers based on Voigt rotation}
\title{Stroboscopic microwave spectroscopy of Voigt based optically pumped magnetometers}
\author{Hans Marin Florez$^{1}$}
\author{Tadas Pyragius$^{2,3}$}
\author{Thomas Fernholz$^{2}$}

% \author{Thomas Fernholz$^{1}$}
%\email{thomas.fernholz@nottingham.ac.uk}
%\email{thomas.fernholz@nottingham.ac.uk}
\affiliation{$^{1}$Instituto de F\'{\i}sica, Universidade de S\~ao Paulo, 05315-970 S\~ao Paulo, SP-Brazil\\
$^{2}$School of Physics \& Astronomy, University of Nottingham, University Park, Nottingham NG7 2RD, UK \\
$^3$ Tokamak Energy Ltd, Milton Park, Oxfordshire OX14 4SD, UK
}

\date{\today}% It is always \today, today,

\begin{abstract}
We present results of stroboscopic microwave spectroscopy of radio-frequency dressed optically pumped magnetometer. Interaction between radio-frequency dressed atoms and a synchronously pulsed microwave field followed by Voigt effect-based optical probing allows us to perform partial state tomography and assess the efficiency of the state preparation process. To theoretically describe the system, we solve the dynamical equation of the density matrix employing Floquet expansion. Our theoretical results are in good agreement with experimental measurements over a wide range of parameters and pumping conditions. Finally, the theoretical and experimental analysis presented in this work can be generalised to other systems involving complex state preparation techniques.
\end{abstract}

\pacs{Valid PACS appear here}% PACS, the Physics and Astronomy
\maketitle

% ----------------------------------------
% ----------------------------------------
\section{\label{sec:intro}Introduction}
% Atomic vapour based optically pumped magnetometers (OPMs) have become the state of the art for highly sensitive magnetic field detection. Since then, applications in several areas have been developed, ranging from fundamental physics, searching for electric dipole moment (EDM), going through geophysical and space magnetometry, up to medicine, such as magneto-encephalography (MEG) and magneto-cardiography.
% Different types of OPMs have shown sensitivity of fT$/\sqrt{\mathrm{Hz}}$, like spin-exchange relaxation-free (SERF) magnetometers
% and radio-frequency excited spin with $M_x$ and $M_z$ magnetometers, which relies in a linear atomic response, and  modulated light magnetometers producing a nonlinear magneto-optical rotation (NMOR), which is based on a nonlinear optical response of the atoms.
%\begin{itemize}
   
 %   \item we need to show the contribution of the density matrix in terms of its matrix elements for the dispersive voigt effect measurement. Since in the Floquet paper we considered the probaility surfaces it would be nice to related them to the multipole moments which contribute to the birefringence expressed in terms of the density matrix elements ans the multipole moments. The exact calculations of this are in my thesis.
 %   \item the current hamiltonian does not include spin-field interaction as well as the equations of motion in the pump-mw-probe regimes (see p93-95 of my thesis).
  %  \item world limit to 4 pages
%\end{itemize}

Atomic vapor based optically pumped magnetometers (OPMs)~\cite{OPM1,OPM2} have become the state-of-the-art devices for highly sensitive magnetic field detection reaching fT$/\sqrt{\mathrm{Hz}}$ sensitivities and beyond. During the last decade OPMs found applications across many domains, from fundamental physics, e.g.\ in the search for electric dipole moments (EDM)~\cite{edm1,edm2}, in geophysical and space magnetometry, and  in medical applications, such as magneto-encephalography (MEG)~\cite{meg1,meg2} and magneto-cardiography~\cite{mcg0,mcg1,mcg2}. 
%The sensitivity depends on the state preparation efficiency.
A commonly used OPM architecture is based on Faraday rotation which further exploits the spin-exchange relaxation-free (SERF) regime ~\cite{Romalis2002,Romalis2010}. In such setups, the atoms are prepared in a spin polarised state, such that in the presence of a magnetic field their precession can be detected by measuring the Faraday rotation through light matter interaction between the atoms and a probe beam. The state preparation is done by circularly polarized pump laser beams reaching a polarizability $P\approx~1$ with non-thermal atoms or highly dense  samples with quenching gas. In the case of Faraday rotation, the state preparation can be verified by using conventional microwave~(mw) spectroscopy~\cite{Wieman93} with a circularly polarised probe beam.
% We need to put P. Truelein ref
% or alternatively probing the light-matter interaction in the non-linear Zeeman regime in the presence of high magnetic fields \cite{brian_state}.
Alternatively, all optical state preparation detection has been demonstrated in cold atoms, but it is not suitable for atomic vapours at room at room temperature, due to the Doppler broadening~\cite{Wang07}.

% In our previous work, we have demonstrated a three-dimensional vector magnetometer based on radio-frequency (rf) dressed states utilising the Voigt rotation~\cite{Tadas19}. 
A different situation is observed when atoms are driven by radio-frequency (rf) fields like in $Mx-$magnetometers~\cite{Groeger06}, double resonance magnetometer~\cite{Weis06} or more recently, as the rf dressed state utilising the Voigt rotation~\cite{Tadas19}.
% Another approach has been proposed based on radio-frequency (rf) dressed states utilising the Voigt rotation~\cite{Tadas19}.
In a Voigt effect based approach, sensitive field measurements are achieved by preparing an aligned state in the presence of a longitudinal magnetic field and an additional rf dressing field. 
% An aligned state corresponds to an equal statistical mixture of two stretched states. 
Such state can, e.g., be prepared as an equal statistical mixture of two stretched states or as a clock state.
By the same token, the efficacy of the state preparation process of an aligned state can be probed by standard microwave spectroscopy~\cite{Avila87}. However, when atoms are additionally driven by a rf field, the Zeeman levels are dressed and the standard selection rules no longer apply. This gives rise to a rich mw spectrum which significantly complicates the inference of the prepared state~\cite{sinuco19}. Alternatively, one can employ a magneto-optical resonance signal (MORS) to probe the energy level distribution~\cite{brian_state04}. However, this requires driving the system to a non-linear magnetic regime, which would destroy the target state that is aimed to be prepared.
% \sout{In addition, the standard microwave spectroscopy in the non-linear Zeeman regime would not allow for the probing of the state as it is needed for the magnetometer, requires additional stages in the state verfication.}\TP{this makes no sense. So here what Hans is saying that state preparation verification usually requires the change of the physical conditions in which the state is verified e.g. going to high fields. Here our method doesn't require the chanhge of such conditions.}

Inspired by previous work in~\cite{sinuco19}, we propose a method of stroboscopic microwave spectroscopy which enables the probing of the population distributions of the prepared atomic states in a dressed scenario in a linear magnetic regime and via Voigt rotation. Furthermore, we demonstrate how the synchronous detection allows to measure the subtle differences in the state preparation with the use of a repump beam. Building upon our previous theoretical work employing the Floquet expansion on second order moments, we extend our theoretical analysis to the density matrix in order to calculate the microwave spectra~\cite{Floquet21}. Using this technique along with our theoretical model, we are able to determine
our target state and estimate the efficiency of the state preparation process reaching more than %approximately
90\%.  We find that our theoretical description is effective in understanding of the experimental results.

% This part was removed for a letter
The paper is organised as follows. In Section~\ref{Sec:Theory} we study briefly describe the spin dynamics and introduce the Floquet expansion to describe the microwave spectroscopy. Section \ref{Sec:Setup} presents the experimental setup and the experimental sequence used to realize the synchronous microwave spectroscopy as well as the results for the synchronous microwave spectroscopy of a Voigt effect based OPM. Section~\ref{Sec:Results} shows the experimental and theoretical results on the synchronous microwave spectroscopy for Voigt effect based OPMs. Section~\ref{sec:Conclusions} presents our conclusions.
%--------------------------------------------------------
\section{Microwave spectroscopy probed by Voigt rotation \label{Sec:Theory}}

The experiments carried out in this work are based on a Voigt effect OPM using radio-frequency dressed states (see ref.~\cite{Tadas19} for further details). The light-matter interaction for such a system is  described by the Stokes parameter
\begin{align}
\Braket{\hat{S}_z'(t)}&=\Braket{\hat{S}_z(t)}+
G_{F}^{(2)} S_y n_F \Braket{\hat{F}_x^2(t)-\hat{F}_y^2(t)}\label{eq:Voigt},
\end{align}
where $\hat{S}_z=(c/2)(\hat{a}^{\dagger}_{+}\hat{a}_{+}-\hat{a}^{\dagger}_{-}\hat{a}_{-})$ and $\hat{S}_y=(c/2)(i\hat{a}^{\dagger}_{-}\hat{a}_{+}-i\hat{a}^{\dagger}_{+}\hat{a}_{-})$ represent the photon flux differences in a circular and a $45\degree$ polarization basis;  $G_{F}^{(k)}$ is the rank-k coupling strength and $n_F$ are the atoms within the same $F$-manifold state~\cite{Jammi18}. 
% which is proportional to transverse second moments of the spin operators i.e the average of the second order products $\Braket{\hat{F}_i(t)\hat{F}_j(t)}$ with $i,j=x,y$ and $z$.
% In order to describe the dynamics of bi-linear operators of the form $\hat{F}_i^2(t)$, a Heisenberg-Langevin approach is required \cite{Floquet20}. However, when dealing with microwave field interactions additional hyperfine ground state  projection operators are needed. This is somewhat involved. A more convenient approach is to employ a density matrix description which can incorporate such interactions more conveniently at the expense of having a larger Hilbert space. Furthermore, it additionally allows us to conveniently apply the Floquet expansion used in our previous treatment of the Voigt effect magnetometers \cite{Floquet20}.
In order to describe the dynamics of bi-linear operators of the form $\hat{F}_i^2(t)$, we adapt the Floquet expansion implemented for the Heisenberg-Langevin description \cite{Floquet21}, to the density matrix description which can incorporate the microwave interaction more conveniently at the expense of having a larger Hilbert space. 

% \subsection{Dynamics with a microwave magnetic field \label{Sec:Dynamics} }
% \HM{To describe the experimental dynamics}, we consider  $^{87}$Rb atoms with hyperfine ground states $F=1$ and $F=2$, \HM{in which the} bare atom Hamiltonian is given by
% $\hat{H}_0=\sum_F E_F \hat{\mathbb{I}}_F$, 

In order to theoretically describe the experimental dynamics, we consider alkali atoms in the ground state $^n S_{1/2}$, in which the orbital angular momentum is $L=0$, hence one can make an approximation  $\hat{\mathbf{J}}=\hat{\mathbf{L}}+\hat{\mathbf{S}}\approx \hat{\mathbf{S}}$. In particular, (without loss of generality), we consider $^{87}$Rb atoms with hyperfine ground states $F=1$ and $F=2$, in which the bare atom Hamiltonian is given by
$\hat{H}_0=\sum_F E_F \hat{\mathbb{I}}_F$, 
% \begin{align}
% \hat{H}_0=\sum_F E_F \hat{\mathbb{I}}_F,
% \end{align}
where the partial identity operator is defined as 
% \begin{align}
%  \hat{\mathbb{I}}_F=\sum_{m_F=-F}^F |F, m_F\rangle \langle F, m_F|.
% \end{align}
$\hat{\mathbb{I}}_F=\sum_{m_F=-F}^F |F, m_F\rangle \langle F, m_F|.$
In addition, we consider the interaction of the atoms in the ground state with a microwave field \cite{sinuco19}
(fast field) and a radio-frequency field (slow field).

We first consider the interaction with the fast field, neglecting the nuclear magnetic moment since it is much smaller than the electronic one, i.e.\ $g_I\mu_B\ll g_S\mu_B$. Thus, the microwave interaction Hamiltonian reduces to
% \begin{align}
% \hat{H}_{\mathrm{mw}}(t)&=\frac{\mu_B}{\hbar}\left(g_I \hat{\mathbf{I}}+g_J\hat{\mathbf{J}}\right)\cdot \mathbf{B}_{\mathrm{mw}}(t),
% \end{align}
% where $g_I$ and $g_J$ are the nuclear and electronic $g$-factors, respectively, with the corresponding angular momentum operators $\hat{\mathbf{I}}$ and $\hat{\mathbf{J}}$. In particular, for alkali atoms in the ground state $^n S_{1/2}$, the orbital angular momentum is $L=0$, hence one can make an approximation $\hat{\mathbf{J}}=\hat{\mathbf{L}}+\hat{\mathbf{S}}\approx \hat{\mathbf{S}}$.
% Moreover, $g_I\ll g_S$, which allows us to neglect the nuclear term. Thus, the microwave interaction Hamiltonian reduces to
\begin{align}
\hat{H}_{\mathrm{mw}}(t)&=\frac{\mu_Bg_S}{\hbar}\hat{\mathbf{S}}\cdot \mathbf{B}_{\mathrm{mw}}(t),
\end{align}
where the microwave field is described classically in the following form 
\begin{align}
\mathbf{B}_{\mathrm{mw}}(t)&=\tilde{\mathbf{B}}_{\mathrm{mw}}e^{-i\omega_{\mathrm{mw}}t}+\tilde{\mathbf{B}}_{\mathrm{mw}}^*e^{i\omega_{\mathrm{mw}}t},
\end{align}
with polarization $\tilde{\mathbf{B}}_{\mathrm{mw}}$ and frequency $\omega_{\mathrm{mw}}$.
The microwave field can be expressed in the spherical basis as $
\tilde{\mathbf{B}}_{\mathrm{mw}}=\beta_{\mathrm{mw}}^0 \mathbf{e}_0 + \beta_{\mathrm{mw}}^+\mathbf{e}_+  +\beta_{\mathrm{mw}}^-\mathbf{e}_-$,
% \begin{align}
% \tilde{\mathbf{B}}_{\mathrm{mw}}&=\beta_{\mathrm{mw}}^0 \mathbf{e}_0 + \beta_{\mathrm{mw}}^+\mathbf{e}_+  +\beta_{\mathrm{mw}}^-\mathbf{e}_-,
% \end{align}
where $\mathbf{e}_0=\mathbf{e}_z$ and $\mathbf{e}_\pm=\mp(\mathbf{e}_x+i\mathbf{e}_y)/\sqrt{2}$, and the magnetic field amplitudes are $\beta_{\mathrm{mw}}^0=\beta_{\mathrm{mw}_z}$ and 
 $\beta_{\mathrm{mw}}^\pm=\mp(\beta_{\mathrm{mw}_x}\pm i\beta_{\mathrm{mw}_y})$.
In the 
% \HM{Supplementary Material I.A} 
Appendix \ref{Sup:MWDynamics}
we show that the dynamics of the density matrix elements interacting with the microwave field under the rotating wave approximation is given by the Liouville equation $\frac{d\tilde{\rho} }{dt}=\frac{i}{\hbar}[\tilde{\rho},\hat{H}_{\mathrm{mw}}^{\mathrm{eff}}]$, where we have defined $\hat{H}_{\mathrm{mw}}^{\mathrm{eff}}=\hat{H}_{0}^{\mathrm{eff}}+\hat{H}_{S}$ with
% \begin{align}
% \hat{H}_{0}^{\mathrm{eff}}=-\begin{bmatrix}
%     0 & 0 & 0 &0 & 0 & 0 &0 & 0  \\
%     0 & 0 & 0 &0 & 0 & 0 &0 & 0 \\
%     0 & 0 & 0 &0 & 0 & 0 &0 & 0 \\
%     0 & 0 & 0 &\Delta&0 & 0 & 0 &0\\
%     0 & 0 & 0 &0&\Delta & 0 & 0 &0\\
%     0 & 0 & 0 &0&0 & \Delta & 0 &0\\
%     0 & 0 & 0 &0&0 & 0 & \Delta &0\\
%     0 & 0 & 0 &0&0 & 0 & 0 &\Delta\\
% \end{bmatrix},
% \end{align}

\begin{align}
 \hat{H}_{0}^{\mathrm{eff}} =\Delta \sum_{m_2} |2, m_2\rangle \langle 2, m_2|,
\end{align}
%\TF{Matrix representations depend on the chosen basis. The minimum must be to state the basis. But this is just the hyperfine operator $A\hat{\mathbf{S}}\cdot\hat{\mathbf{I}}$ in the $F$-basis (with shifted energy reference). Why do you call it "effective"?}
where $\Delta$ corresponds to the frequency detuning of the microwave with respect to the hyperfine clock transition, whereas the microwave field Hamiltonian can be expressed as
\begin{align}
\hat{H}_{S}&=\Omega_\pi \hat{S}_z +\Omega_{\sigma+} \hat{S}_{\sigma+}+\Omega_{\sigma-} \hat{S}_{\sigma-},
\end{align}
where the new spin operators are defined in eqs.(\ref{eq:Sz}-\ref{eq:Sn}).
The dynamics describe the Rabi oscillations induced by the microwave field coupling the two hyperfine ground states of alkali atoms.

Let us now consider the interaction with the static field $B_\mathrm{dc}$ and the rf dressing field $B_{\mathrm{rf}}$, the slow field interaction is  $\hat{H}_B=(\mu_Bg_F/\hbar)\hat{\mathbf{F}}\cdot\mathbf{B}$, where $\mu_B$ and $g_F$ correspond to the Bohr magnetron and $g_F$-factor, respectively. The total magnetic fields is
\begin{equation}
\mathbf{B}=(B_{\mathrm{rf}} \cos{\omega_{\mathrm{rf}} t} +B_x^{\mathrm{ext}})\mathbf{e}_x+B_y^{\mathrm{ext}}\mathbf{e}_y+(B_{\mathrm{dc}}+B_z^{\mathrm{ext}})\mathbf{e}_z,\label{eq:Blab}
\end{equation}
where $\omega_{\mathrm{rf}}$ corresponds to the frequency of the rf field and we consider the general case in which external fields $\mathbf{B}^{\mathrm{ext}}=(B_x^{\mathrm{ext}},B_y^{\mathrm{ext}},B_z^{\mathrm{ext}})$ are present.
Hence, the total coherent dynamics is given by $\hat{H}_{T}(t)=\hat{H}_{\mathrm{MW}}^{\mathrm{eff}}+\hat{H}_{B}(t)$. Including the pumping rate $\Gamma_p(t)$ which allows the state preparation with an arbitrary time profile and the relaxation rate $\gamma$ due to collisions, the total dynamics can be described by  
% \HM{\ref{S-Sup:rfDynamics}}
% \footnote{The detailed description of the atomic dynamics is presented in Supp.A.}
\begin{align}
\frac{d\tilde{\rho} }{dt}=\frac{i}{\hbar}[\tilde{\rho},\hat{H}_{T}(t)]-\Gamma_p(t)(\tilde{\rho}-\rho_{in}) - \gamma (\tilde{\rho}-\rho_{0}).\label{eq:rhoT_dynamics}
\end{align}
defined in 
% \HM{Supplementary Material I.B.}
Appendix \ref{Sup:rfDynamics}.
Solving the dynamics in a general situation in which the pumping rate and the microwave fields are modulated with an arbitrary time profile at a frequency $\omega_\mathrm{rf}$, is not straightforward. However, by applying the Floquet expansion to the density operator, as in ref.~\cite{Floquet21}, it is possible to find a steady state solution. Considering a harmonic expansion of the density matrix elements as $\rho_{ij}(t)=\sum_n\rho_{ij}^{(n)}(t) e^{in\omega_{\mathrm{rf}} t}$, with $n\in Z$,  it is possible to find the steady state solution of the harmonics  $\rho_{ij}^{(n)}(t)$,  which is detailed described in 
% \HM{Supplementary Material \ref{S-Sup:FloquetExpan}.}
% \HM{Supplementary Material I.C.}
Appendix \ref{Sup:FloquetExpan}.

% \TP{here the text appears to abruptly stop. The next step should be to describe rf dressed spin-field interaction Hamiltonian and then finally combine it with the effective Hamiltonian.}

 \begin{figure}[t!]
\begin{overpic}[width=0.4\textwidth]{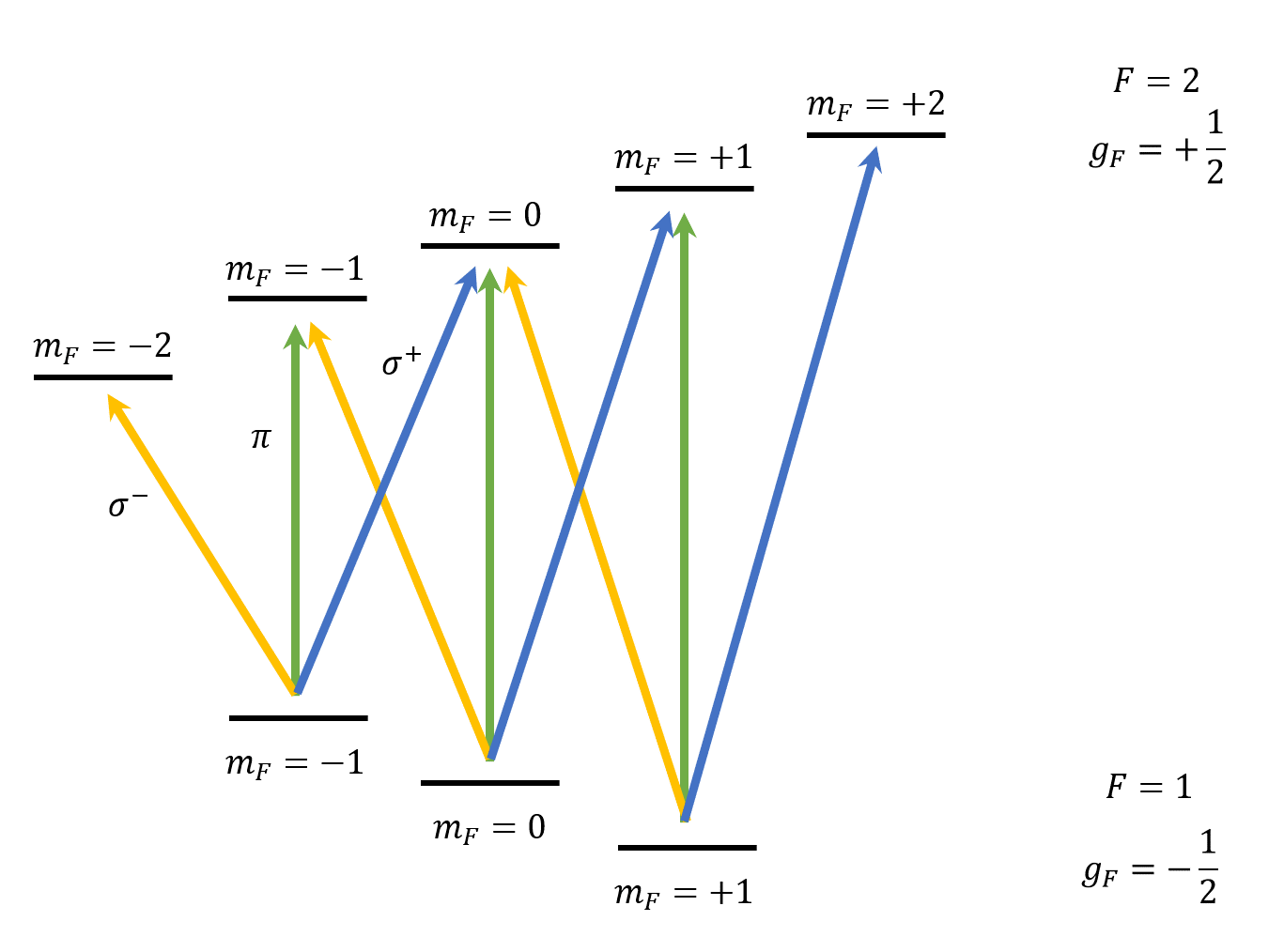}
% \begin{overpic}[width=0.5\textwidth]{mw_transitions.png}
\put(-10,65){(a)}
\end{overpic}
\begin{overpic}[width=0.5\textwidth]{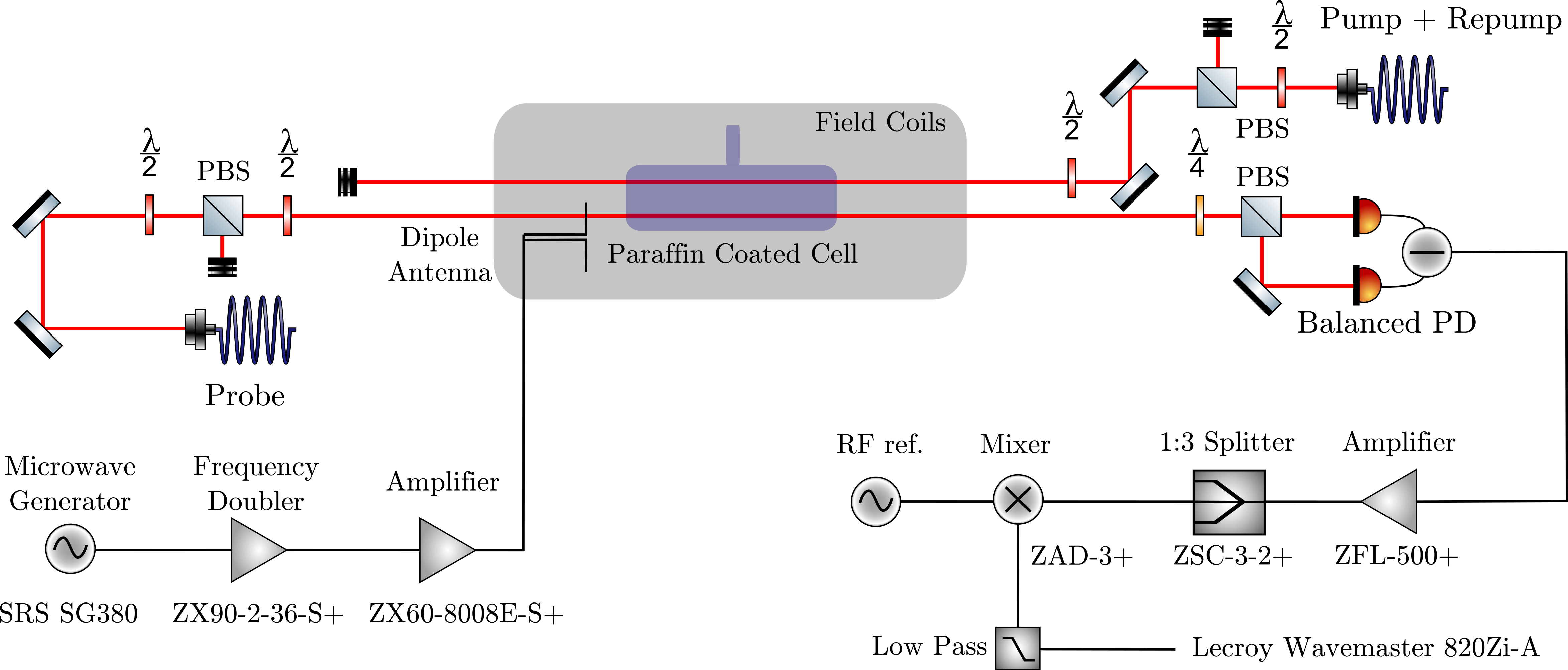}
\put(0,40){(b)}
\end{overpic}
\caption{(a) Bare Zeeman states of $^{87}$Rb and polarisations of mw transitions. (b) Sketch of the experimental setup.}
\label{fig:setup}
\end{figure} 
%  Interaction with the magnetic field
 
\section{Experimental setup\label{Sec:Setup}}

Our magnetically unshielded experimental setup is depicted in Fig.~\ref{fig:setup} and  its detailed description is presented in the
Appendix \ref{Sup:exp_setup}.
% Supplementary Material II.
It follows closely the shielded version of the experimental setup described in ref.~\cite{Tadas19}. A paraffin coated $^{87}$Rb enriched vapour cell 
% {\color{black}of diameter} $d=26$~mm and {\color{black}length} $l=75$~mm 
at room temperature.
% The atomic state is dressed with a radio-frequency field which is generated by cosine-theta coil along x-axis. 
The atoms are dressed with a $\omega_\mathrm{rf}=2\pi\times~90$~kHz rf field along the $x$ direction and coupled to a static field in the longitudinal direction. Three pairs of Helmholtz coils are used to actively compensate and stabilise the external magnetic field. The atomic state is prepared by a pump laser beam tuned to the $F=2$ to $F'=1$ transition of the D1 line. The atoms are stroboscopically pumped by modulating the pump amplitude with 10\% duty cycle in phase with the rf field.

To probe the state precession, we employ the Voigt rotation for non-destructive measurements in hyperfine ground states $F=1$ and $F=2$. 
% To do so, the atoms are probed by a $45^\circ$ polarized laser beam relative to the pump polarization and tuned $-500$~MHz with respect to the $F=2$ to $F'=1$ transition and $F=1$ to $F'=1$ transition of the D1 line. After the interaction with the atoms, the light passes through a quarter waveplate and a polarizing beam splitter, which allows us to measure the Voigt rotation of the light polarization i.e. the ellipticity induced by the atoms. 
The light is detected on a balanced photodetector  where the detected signal $u(t)=g_{el} S_z(t)$ is proportional to the ellipticity in eq.~(\ref{eq:Voigt}) and the electronic gain $g_{el}$. As it was shown in ref.~\cite{Tadas19}, the ellipticity for the Voigt rotation produces a signal at the first and second harmonic of the radio-frequency dressing field such that $u(t)=m_0+m_1e^{i\omega_\mathrm{rf}t}+m_2e^{2i \omega_\mathrm{rf}t}+c.c$. This output signal is demodulated at the second harmonic, from which we can extract its mode amplitude $m_2$.
% Description of tuning the 2f resonance 
The microwave spectroscopy is performed with the magnetometer tuned on resonance, $B_z=\hbar\omega_\mathrm{rf}/\mu_Bg_F$ using the second RF harmonic in the Voigt rotation signal, cancelling the presence of any transverse field, i.e.\ zeroing the first harmonic~\cite{Tadas19}.
%Description of the antena
% The microwave field is generated using an rf generator (SRS SG380) with a frequency doubler (Minicircuits ZX90-2-36-S+) and an additional amplifier. The signal is coupled into a half-wave dipole antenna $L=\lambda/2$ where $L$ is the conductor length and $\lambda$ corresponds to $\Ket{F=1}\rightarrow\Ket{F=2}$ hyperfine microwave transition.
The direction of the dipole antenna is transversal to the light propagation. 
% The experimental and theoretical sequences to obtain the mw spectra are shown in Fig.~\ref{fig:cone_cw_mw} and Fig.~\ref{fig:cone_pulsed_mw}.
% \HM{We need to put the figure of the experimental setup}.
%--------------------------------------------------------
% \section{Results\label{Sec:Results}}

% In what follows we report on the results of the rf dressed microwave spectroscopy in two distinct microwave interaction regimes. The first regime is where the atoms are continuously interacting with the microwave field during the mw stage. The second regime consists of the atoms interacting with a stroboscopically pulsed mw in phase with the rf dressing field. During the interaction with the microwave field all optical interactions are switched off such that the atomic dynamics are solely governed by the atom-rf-mw couplings, see Fig.~\ref{fig:cone_cw_mw}.

%--------------------------------------------------------
% \subsection{Microwave CW\label{Sec:MWCW}}
% A typical bare micro-wave spectrum for the two hyperfine ground states by measuring the transmission of a probe beam is composed by seven peaks coupling  the  longitudinal and transverse microwave polarizations, as it is shown in Fig.~\ref{fig:setup}~(b).

\section{Results \label{Sec:Results}}

\begin{figure}[t!]
\begin{center}
\begin{overpic}[width=8.6cm]{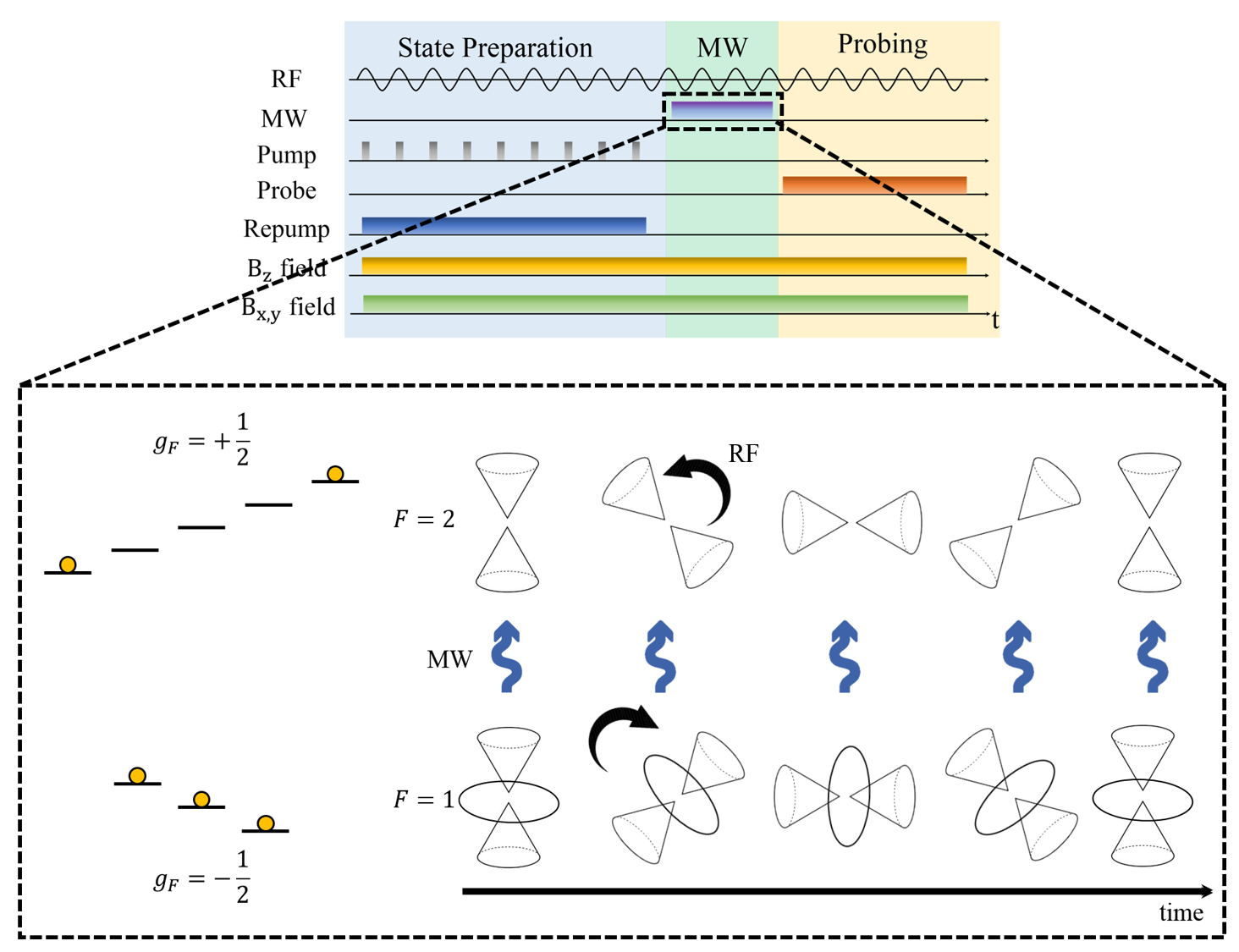}
\put(0,75){(a)}
\put(3,40){(b)}
\end{overpic}
\end{center}
\caption{Illustration of the experimental sequence and the microwave coupling of the ground state manifolds in the continuous-wave (CW) model. The sequence consists of three separate stages; state preparation, microwave interaction and probing. The states in each manifold rotate in opposite directions due to the different $g_F$-factors. The microwave radiation transfers the state populations between the two energy levels at different cone orientations during the evolution.}
\label{fig:cone_cw_mw}
\end{figure}

We first perform a continuous microwave spectroscopy during relaxation dynamics of the prepared state, as it is shown in Fig.~\ref{fig:cone_cw_mw}~(a).
For a microwave field composed of $\sigma^{\pm}-$ and $\pi-$ polarizations applied to $F=1\rightarrow F=2$ in the presence of an external static magnetic field, there are a total of 9 possible microwave transitions, 4 of which are degenerate, see Fig.~\ref{fig:setup}~(a). This results in 7 distinct resonance peaks in the undressed microwave spectrum when the microwave field frequency is scanned (assuming a thermal state in each manifold), separated by $\Delta\omega=\mu_Bg_FB_z/\hbar$. If the state is polarised, e.g. we have a stretched state, then only one transition will be observed. If, as in our case, an equal statistical mixture $\Ket{F=2,m_F=\pm 2}$ is prepared, then the two extreme transitions should be observed.

\begin{figure}[t!]
\begin{overpic}[width=0.5\textwidth]{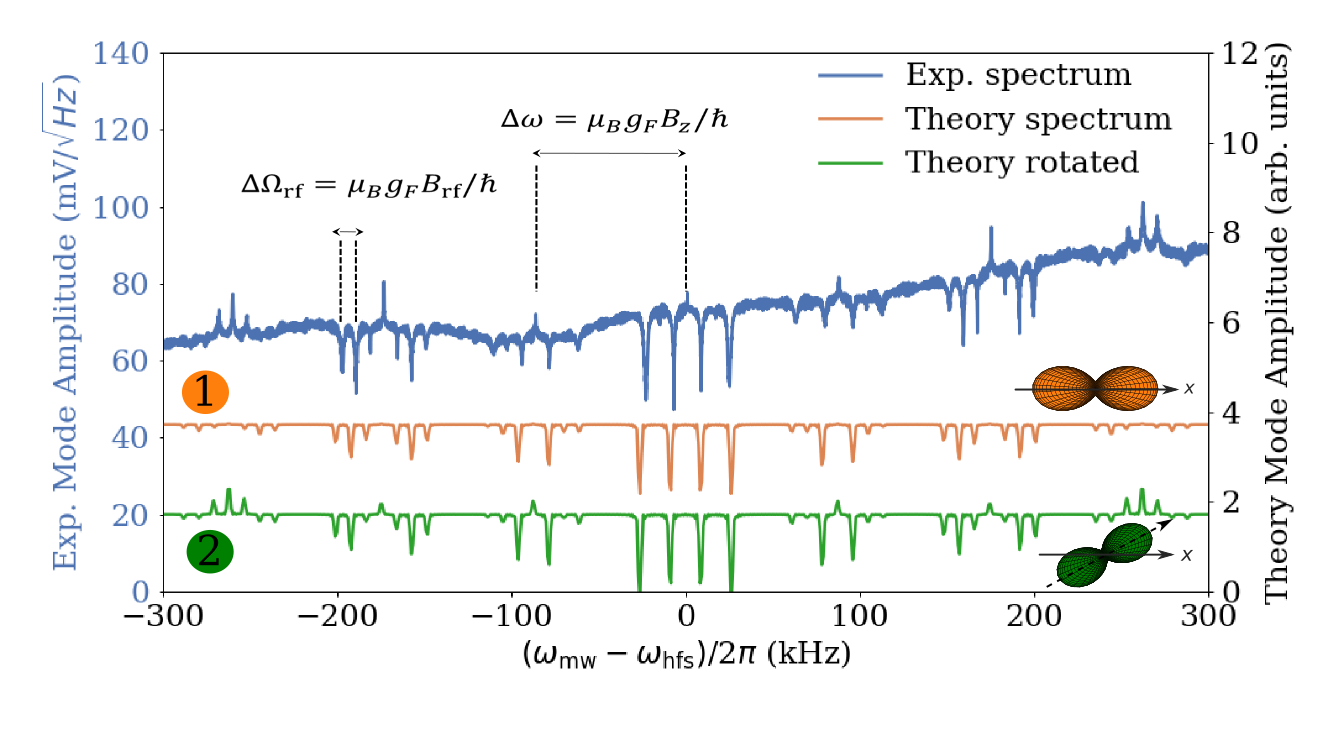}
\put(0,53){(a)}
\end{overpic}
\begin{overpic}[width=0.495\textwidth]{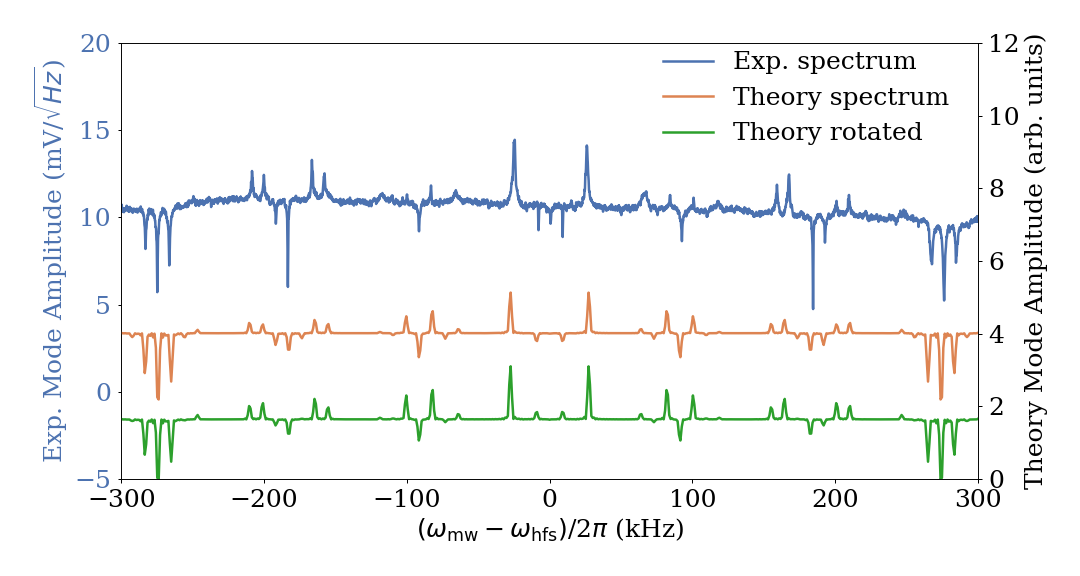}
\put(0,50){(b)}
\end{overpic}
\caption{Experimental and theoretical dressed microwave spectra of the $\mathrm{Re}(m_2)$ mode amplitude as a function of the microwave frequency. Here the measured and calculated spectra are obtained based on the sequence in Fig.~\ref{fig:cone_cw_mw}. (a) Probe $F=2$, mw CW, repump ON. (b) Probe $F=2$, mw CW, repump OFF. Here the theoretical mw is applied in a CW mode over 20 rf periods, $T=2\pi/\omega_\mathrm{rf}$. In both cases we considered an effective rotation of $\sim 35^\circ$.}
\label{fig:CW_MW_spectrum_F2}
\end{figure}

% The transitions are known as group transitions. 
Now, in the case of rf dressed states, the spectrum includes new transitions in the dressed basis. % Figures (\ref{fig:CW_MW_spectrum_F1}) and (\ref{fig:CW_MW_spectrum_F2})
% Figure \ref{fig:CW_MW_spectrum_F1}~(a) and (b)
% show the theoretical and experimental rf dressed microwave spectra of the Re($m_2$) demodulated signal amplitude for $F=2$ and $F=1$ manifolds, respectively, with an active repump. 
Figure \ref{fig:CW_MW_spectrum_F2}
shows theoretical and experimental rf-dressed microwave spectra of the Re($m_2$) demodulated signal amplitude for $F=2$ manifolds for two differently prepared states.
As described in \cite{sinuco19}, the general structure of these spectra comprises seven groups of up to seven lines. Different groups are addressed by different mw polarizations, but the complete set arises here due to the inhomogeneous field distribution from the dipole antenna. The group structure resembles that of bare transitions, alternating between $\pi$ and $\sigma$-polarisations, with a spacing given the static field strength. Lines within a group are spaced by $\sim\sqrt{\Omega_{\mathrm{rf}}^2 + \Delta^2}$. Since in our case the rf is in resonance with the Zeeman levels, $\Delta=0$, the line spacing measures the rf amplitude, $\Omega_\mathrm{rf}$. This can be used
to obtain absolute field values for both the static and rf fields which provides a method for absolute ac and dc field calibrations.

% The first, third, fifth and seventh groups are associated with the sigma transitions, transverse to the pump polarization, whereas
% the second, fourth and sixth groups are associated with the longitudinal polarization. The fourth group (middle) corresponds to the clock transition in bare spectrum terms. 

% It is worth noting that the mw spectrum for the $F=1$ manifold yields a zero mode amplitude, Re($m_2$), whilst the mw spectrum for $F=2$ probed manifold possesses a non-zero mode amplitude indicating non-zero birefringence. In the case of $F=1$ probing, the state preparation process with or without repump addressing $F=1\rightarrow F'$ optical transition will result in a completely depopulated $F=1$ manifold or a manifold containing a thermal state. In both such cases the birefringence of the two systems when the $F=1$ manifold is probed is zero which results in zero value of the mode amplitude Re($m_2$). Consequently, an interaction between the $F=1$ and $F=2$ manifolds via microwave coupling induces transitions thereby introducing an imbalance in the state populations which results in non-zero birefringence. 

Fig.~\ref{fig:CW_MW_spectrum_F2}~(a) shows data with repump light, intended to prepare a mixture of the two rf-dressed, fully stretched states. Here, population that leaks into the $F=1$ manifold is redistributed into the $F=2$ manifold, where it interacts with the pump beam. This results in a zero birefringence signal in the $F=1$ manifold and non-zero birefringence in the $F=2$ manifold, which is seen as a large offset in the data.
% In the case of $F=1$ probing, the state preparation process with or without repump addressing $F=1\rightarrow F'$ optical transition will result in a completely depopulated $F=1$ manifold or a manifold containing a thermal state. 
%  In both such cases the birefringence of the two systems when the $F=1$ manifold is probed is zero which results in zero value of the mode amplitude Re($m_2$). 
The prepared state is close to maximal birefringence and any coupling of the $F=1$ and $F=2$ manifolds by resonant microwave transitions alters state populations and becomes visible mostly as spectral dips.    
% Comparing theoretical and experimental spectrum
However, the rf-dressed microwave spectrum possesses a complex and difficult to interpret structure compared to a bare spectrum.
This is because the rf radiation provides additional coupling within the Zeeman states corresponding to coherences in the density matrix which contain information about the correlations between any two given populations. %The picture is made more complicated by the fact that our atomic spins are dressed with a radio-frequency field which additionally couples the hyperfine sublevels, see Fig.~\ref{fig:cone_cw_mw} for a schematic illustration. 
If the mw field is turned on all the time, then the evolution of the mw field and the state is averaged, i.e.\ different relative orientations between the cones end up contributing to the overall spectrum, as it is shown in Fig.~\ref{fig:cone_cw_mw}~(b). 
The coupling of the mw field from one manifold to the other probes the orientation of those two cones relative to each other as the mw field frequency is continuously scanned.
This allows the probing of all the possible transitions resulting in a dense forest of lines, rendering the standard selection rules obsolete. For example, coupling $\Ket{F=1, m_F=-1}$ and $\Ket{F=2, m_F=2}$ is possible. Hence, addition of the rf dressing field gives rise to partially degenerate multi-photon transitions, enabling the coupling between any two given states~\cite{sinuco19}.
Now, let us consider the situation in Fig.~\ref{fig:CW_MW_spectrum_F2}~(b), which shows microwave spectra of the OPM without the repump, for the Voigt rotation in $F=2$. 
If the repump is turned off a fraction of the population is  pumped into $F=1$ and a another fraction remains in $F=2$, where the pump beam drives the atoms to an aligned state with lower efficiency.
% Figure \ref{fig:CW_MW_spectrum_F2}~(b) and (b) shows the theoretical and experimental microwave spectra of the OPM without the repump, for the Voigt rotation in $F=1$ and $F=2$, respectively~\footnote{For the theoretical calculation we employ Bloch equations for the D1 line of $^{87}$Rb  and calculate the steady state solution of the required situation with no repump. From the total density matrix we extract the corresponding ground states density matrix and use it as the input state for our theoretical model.}. 
Unlike in the previous case, the $F=1$ manifold is now populated, but is in a near-thermal state with vanishing birefringence, which strongly diminishes the $\mathrm{Re}(m_2)$ offset amplitude. Nevertheless, here the microwave coupling between the two ground-state manifolds enables a 2-way population exchange, which is distinct to the previous case. The mw transitions may now change birefringence in either sense.
In Appendix \ref{supl:F1_spectrum}, we also present mw spectra for $F=1$ comparing the cases with and without repump light showing that this key difference enables one to distinguish between a completely empty and thermally populated manifolds which otherwise posses identical zero birefringence.

For the further discussion, we focus on the $F=2$ manifold where we prepare the atomic state. In this case the theoretical mw spectrum in orange displays similar structure as the experimental one, with some differences. In particular, for the case of active  repump in Fig.~\ref{fig:CW_MW_spectrum_F2}~(a), curve (1) shows that the two extreme groups are clearly different from the the experimental observations, whereas for the case of no repump, the same groups are properly described. Since in the first case the $F=2$ population is higher than in the second case, propagation effects can be present during the state preparation process. Hence, if one assumes an effective rotation of the aligned state around the static field $z$-axis due to the propagation effects, curve (2) is now in agreement with experimental observations. This is not the case for spectrum in Fig.~\ref{fig:CW_MW_spectrum_F2}~(b), where applying the same rotation effect does not change the extreme group profiles but instead modifies the centre group. This indicates that for no repump, the atomic $F=2$ population is low compared to the case with active repump, and no propagation effects have to be considered.

Although, from the theoretical model the prepared state used to obtain the spectrum corresponds to an aligned state, in both cases of Fig.~\ref{fig:CW_MW_spectrum_F2},  it is not straightforward to infer the population distributions in the rf dressed picture, i.e.\ the quasi-energy eigenstates. As a result, the need for an alternative probing scheme is required. Here we propose a rf-synchronised stroboscopic mw probing scheme in order to map the population distributions in the presence of an rf dressing field.

\begin{figure}[t!]
\begin{overpic}[width=0.5\textwidth]{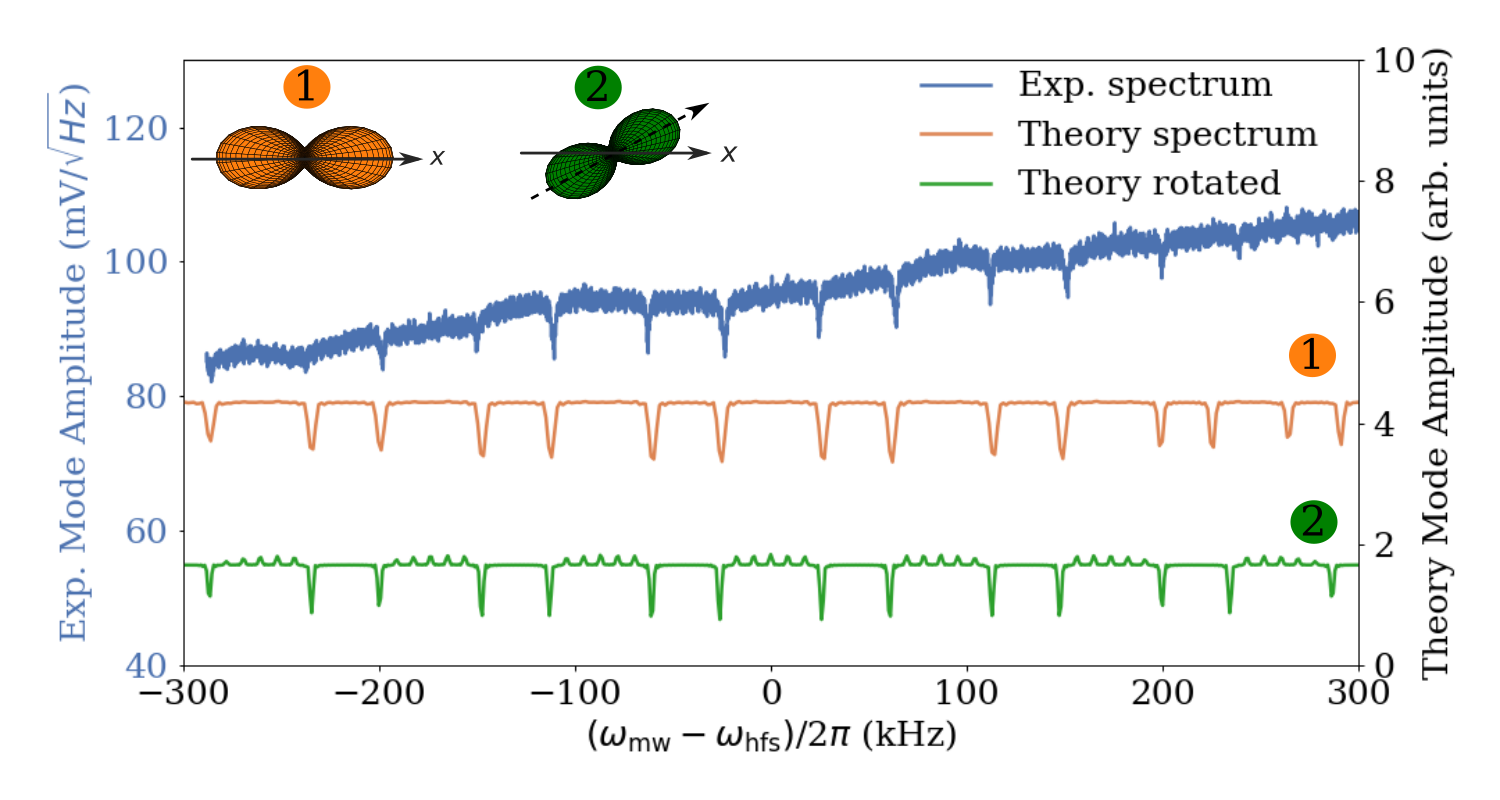}
\put(0,50){(a)}
\end{overpic}
\begin{overpic}[width=0.5\textwidth]{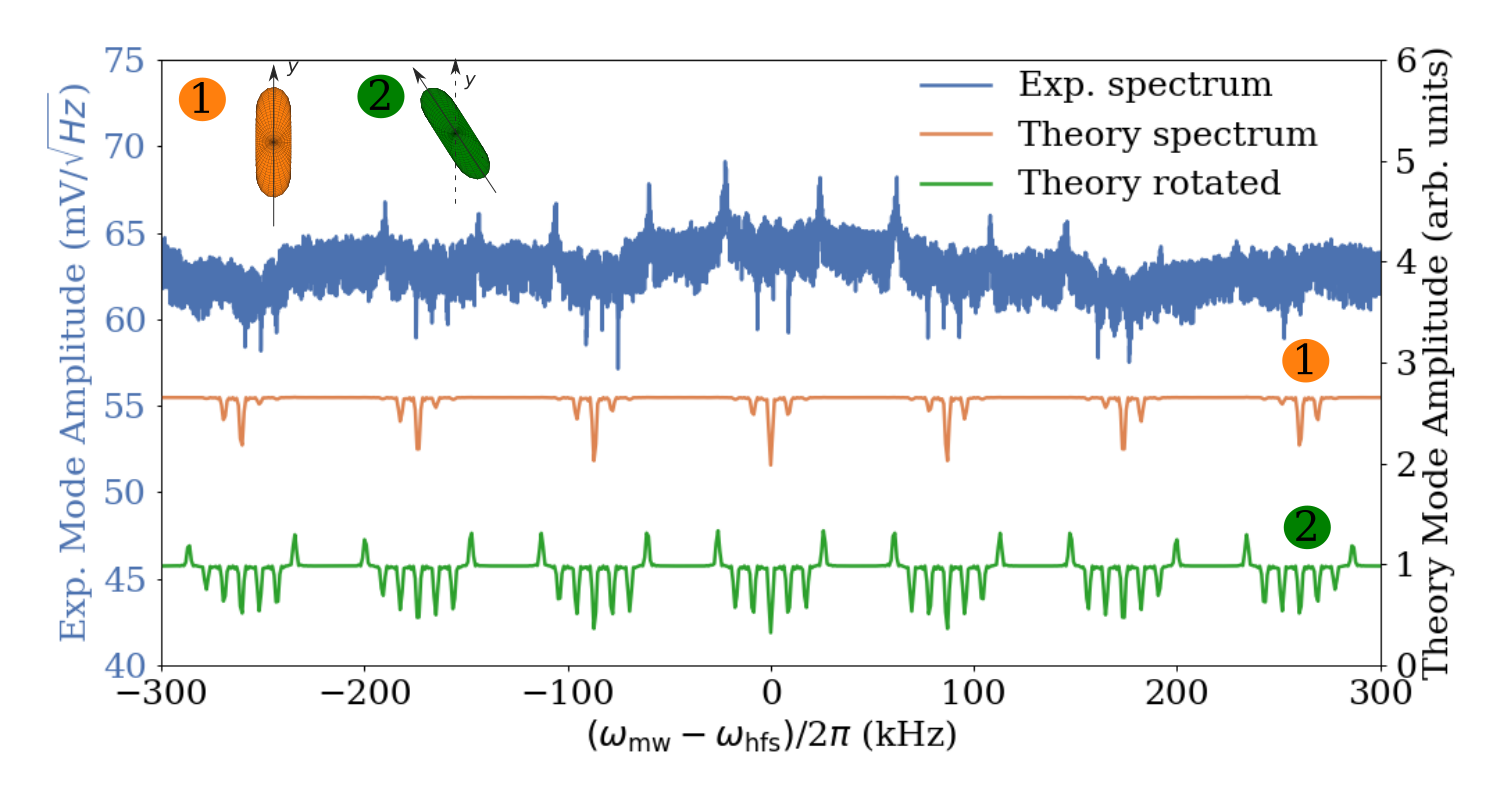}
\put(0,50){(b)}
\end{overpic}
\caption{Experimental and theoretical dressed microwave spectra of the $\mathrm{Re}(m_2)$ mode amplitude as a function of the microwave detuning with respect to the clock transition.
% frequency (for simplicity we plot the detuning on the x-axis, 0~kHz frequency corresponds to the clock transition of $^{87}$Rb). 
Here, the measured and calculated spectra are obtained based on the sequence in Fig.~\ref{fig:cone_pulsed_mw} where the mw is pulsed. (a) Extremal aligned state with repump ON. (b) Clock state with repump ON. In both cases we considered an effective rotation of $\sim 35^\circ$.}
\label{fig:PL_MW_spectrum_F2}
\end{figure}

% \subsection{Synchronous Microwave\label{Sec:SyncMW}}
Figure~\ref{fig:PL_MW_spectrum_F2}~(a) shows the results of pulsed microwave spectrum  of Re$(m_2)$ probing the $F = 2$ ground state manifold, for the situation analysed in Fig.~\ref{fig:CW_MW_spectrum_F2} when the repump is active.
%of the Re$(m_2)$ quadrature of the Voigt effect.
Here the microwave field is pulsed with the same duty cycle of 10\% and in phase with the radio-frequency dressing field. The experimental stroboscopic microwave spectrum shows %only 
mainly two peaks per group which lie on the two extreme edges. However, some of the manifold populations leak into the inner levels.
On the other hand, the theoretical calculation shows that, when the system is prepared in an equal statistical mixture of the stretched states and then probed stroboscopically using mw radiation in phase with the dressing rf field, the atomic populations perfectly distributed into two extrema peaks in the dressed microwave spectrum repeated over the 7 groups~\footnote{The multiple reproduction of the single mw dressed spectrum is due to the harmonics generated by the square modulation of the mw field.}. This is expected for an extremal aligned state, in which the majority of the populations reside in the Zeeman sublevels of the $|F = 2,m_F = \pm 2\rangle$ mixed state along the x-axis. Now, the inner peaks detected on the mw scan can be described by  taking into account the propagation effect as in Fig.~\ref{fig:CW_MW_spectrum_F2}. Notice that if we consider some effective rotation due to the propagation effects in the theoretical calculation, the resultant theoretical spectrum matches the experimental observations, showing the presence of the inner peaks. 
% This indicates that  our state preparation process is efficient with approximately 99\% of the atoms ending up in the target state. 
This indicates that without propagation effects  our state preparation process can in principle reach near perfect efficiency with the atoms ending up in the target state.
However, when the propagation effects are considered, the aligned state is rotated, thus reducing the preparation efficiency to $\sim 70\%$ of the mixture of the two stretched states.

\begin{figure}[t!]
\begin{center}
\begin{overpic}[width=0.5\textwidth]{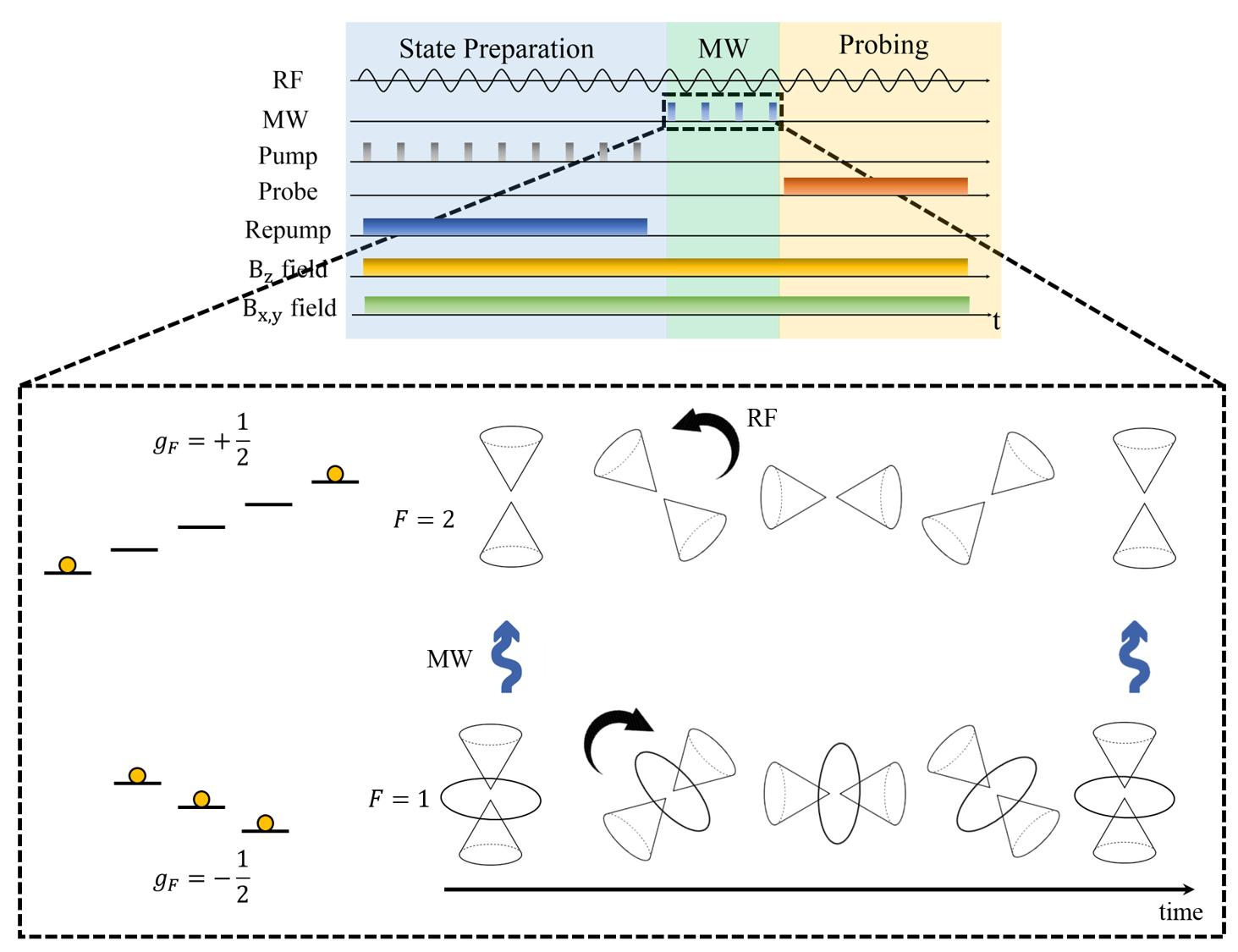} \end{overpic}
\end{center}
\caption{Illustration of the experimental sequence and the the microwave coupling of the ground state manifolds during stroboscopic microwave probing. Here the states in each manifold are probed only at a specific time (stroboscopic microwave interaction) which means that the population transfer between the two levels happens only at a certain orientation between the two cones. The microwave pulses are in phase with the rf field.}
\label{fig:cone_pulsed_mw}
\end{figure}

This stroboscopic mw spectrum measurement can be described as probing the states at a certain orientation in their evolution as they precess around the static and rf fields, see Fig.~\ref{fig:cone_pulsed_mw}. As a result, when the mw field is pulsed and its frequency scanned, only a certain part of the cone configuration is probed and so the spectrum acquires only the shape of the corresponding bare spectrum, or rather, the spectrum is obtained based on a single orientation of the cones. 
% Imagine this as a certain effective angle between the cones that happens at a certain time window and each pulse probes only at that specific cone orientation - this is stroboscopic probing. The bare spectrum for a short mw pulse reveals the information of relative population sizes for each state. 

% Therefore, by playing with pump frequency, polarisation and the repump laser one wants to make sure that the bare spectrum yields populations which are in the extremes - implying that most of the atoms are prepared in the aligned state,% see Fig.~\ref{fig:pulsed_mw_theory} for results. 

% The theoretical calculations show that when the system is prepared in an equal statistical mixture of the stretched states and then is probed stroboscopically using mw radiation in phase with the dressing rf-field, the the state populations appear as the two extrema peaks in the dressed microwave spectrum repeated over the 7 groups. In a fully undressed scenario, such a state would produce the same features, but only in the two extreme groups. 
% As a result, this stroboscopic microwave probing can be used as a smoking gun to infer the prepared state and its corresponding population levels. In addition, the phase of the microwave pulse is changed, then we end up probing different orientations during the state evolution. If we probe the spectra at different rf-phases and then add them, this would yield the mw spectrum when the mw is in cw mode.

To further test the efficacy of our theoretical model, we consider the situation where the the pump couples the $F=2\rightarrow F'=2$ transition which prepares the dressed "clock" state, $|F=2,m=0\rangle_x$.
% as it is shown in Fig.~\ref{fig:PL_MW_spectrum_F2}~(b).
In this case, one can notice that the state preparation does not just produce the clock state, %$|F = 2,m_F =0\rangle$,
but a distribution within the manifold. Once again, the theoretical model shows that if the input state corresponds to a clock state in the $x$-direction, the stroboscopic spectrum should be mainly determined by the clock state. However, when considering the propagation effects, the model predicts the population distribution to the one observed experimentally.
In addition to these observations, we show in the Appendix \ref{supl:F1_spectrum} that the experimental and theoretical mw spectra in the CW and stroboscopic limits are also in agreement when probing the $F=1$ manifold, when the propagation effects are considered. 

Details of our theoretical model are given Appendix \ref{Sup:Dynamics}. Extending the model in Ref.~\cite{sinuco19}, which describes isolated atomic systems, this model describes an atomic medium with pumping and relaxation rates. So far we have just discussed the mw spectra when no transverse static fields are applied, focusing on the high amplitude of the second harmonic. However, the model can predict also the CW and the pulsed mw spectrum for the first and higher harmonics in the presence of transverse fields, contained in the magnetic interaction $\hat{H}_B(t)$ of the total Hamiltonian in Eq.~(\ref{eq:rhoT_dynamics}). Therefore, the model is quite robust in being able to describe the mw spectroscopy of the spins in different scenarios.

In order to experimentally overcome the propagation effects, a shorter cell can be used to reduce the back action of the pump for relatively high atomic populations. This would induce smaller effective rotations in the beam polarisation. Nevertheless, this theoretical model allows us to have satisfactory predictions of the expected stroboscopic spectra and their relation to the prepared input state. 
% What is the input state population in x?
% This state has a similar surface profile to the ideal aligned state, with some subtle differences. 
% The theoretical model shows the same kind of population leakage observed in the experimental data, which indicates that this state is slightly different from an ideal aligned state. \HM{Approximately 45\% of the atomic population of the manifold is prepared in both stretched states.} 

%i.e. the purity of the final state is reduced with respect of an aligned state \TP{we have to be careful here when we use the term pure to describe our quantum states. The context here is not clear. We produce an equal statistical mixture composed of pure states, but the state we produce itself is not pure. Are you saying here that the purity is related to some residual populations?}.

\begin{figure}[t!]
\begin{overpic}[width=0.5\textwidth]{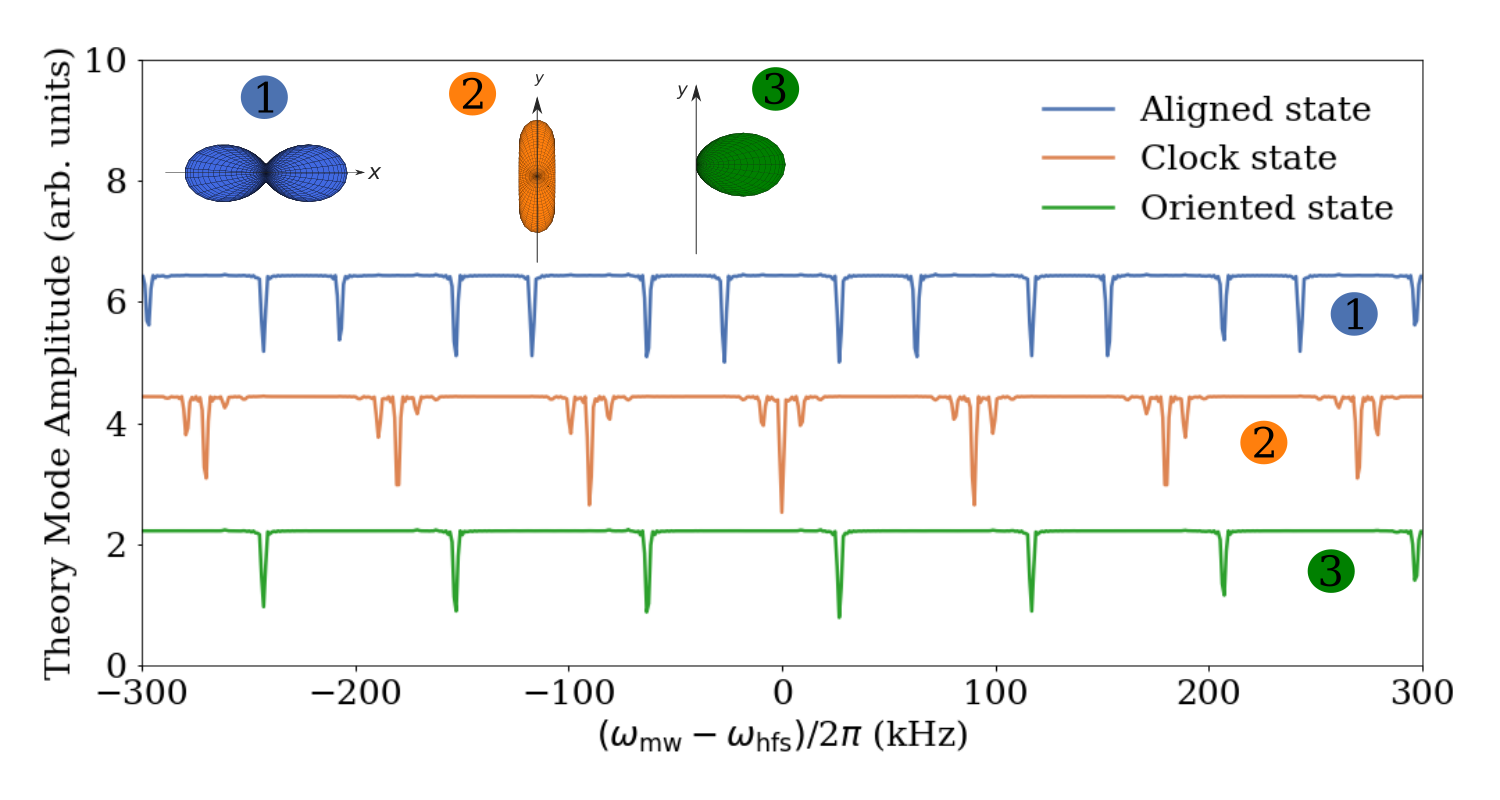}
\end{overpic}
\caption{Theoretical dressed microwave spectra of the $\mathrm{Re}(m_2)$ mode amplitude as a function of the microwave frequency detuning. Here the measured and calculated spectra are obtained based on the sequence in Fig.~\ref{fig:cone_pulsed_mw} where the mw interaction is stroboscopic and no propagation effect is considered.  }
\label{fig:PL_MW_spectrum_F2_theo}
\end{figure}

An additional comparison can be made with the help of the theoretical model. A synchronously dressed oriented state in an rf magnetic field yields atomically modulated birefringence at twice the modulation frequency of the dressing field analogous to atomically modulated birefringence observed in clock or aligned states~\cite{Floquet21}. Figure \ref{fig:PL_MW_spectrum_F2_theo} shows the pulsed mw spectra for the three distinct states considering the ideal case, where propagation induced rotation effects are negligible. 
Notice that the stroboscopic spectroscopy can give access to atomic population distributions according to each input state e.g. two extreme peaks for the mixture $|F=2, m_F=\pm2\rangle$, the centre peak for the clock state $|F=2, m_F=0\rangle$ and one extreme peak for the oriented state $|F=2, m_F=2\rangle$. This indicates that, under negligible propagation effects, the stroboscopic mw spectroscopy can be used to measure the atomic population distributions when atoms are dressed by rf-fields.
 
% ----------------------------------------
% ----------------------------------------
\section{\label{sec:Conclusions}Conclusions}

We have presented a novel method of microwave spectroscopy of radio frequency dressed states applied to a Voigt effect based optically pumped magnetometer. In the case of continuous microwave interaction, we have shown the violation of standard selection rules manifesting in multiple microwave transitions among the rf dressed states in the hyperfine ground state manifold. We have demonstrated that it is possible to overcome the complex interpretation of the mw spectrum in CW configuration, by changing the mw interaction profile to a rf synchronised stroboscopic microwave pulsing. 
Applying this technique, we have shown that one can model the process of an efficient state preparation of an extremal aligned, clock and oriented state driven by a combination rf fields and as well as optical pump and repump beams. 
% Otherwise, efficiency of a linearly polarized pump for preparing an aligned state is reduced showing population leakage to other Zeeman sub-levels, which for the first time, allows us to study subtle differences in the state preparation process of aligned states.
Furthermore, the method of state probing outline enables us to distinguish between thermal an empty manifold populations which otherwise posses zero birefringence.
Finally, our proposed approach can be easily extended to radio-frequency dressed Faraday measurements to map the state population distributions, as well as any other dynamical state preparation processes.
%\HM{However, this detection scheme is suitable of propagation effects which can be avoided by employing short vapour cells.}
This work paves the way towards a new form of dressed state detection and mapping.

% ----------------------------------------
% ----------------------------------------
\section{Acknowledgements}
This work was funded by Engineering and Physical Sciences
Research Council (EP/M013294/1) and by  Grant  No.  2018/03155-9  S\~ao  Paulo  Research  Foundation
(FAPESP).
% ----------------------------------------
% ----------------------------------------
% ----------------------------------------

\appendix

\section{Theoretical description of the spin dynamics with time dependent magnetic fields \label{Sup:Dynamics} }
In this section we present the theoretical description of the evolution of an atomic spin interacting with microwave and radio frequency magnetic fields with decoherence effects governed by Langevin dynamics.

\subsection{Spin dynamics with microwave fields \label{Sup:MWDynamics} }
Consider $^{87}$Rb atoms with hyperfine ground states $F=1$ and $F=2$. The free Hamiltonian is given by
\begin{align}
\hat{H}_0=\sum_F E_F \hat{\mathbb{I}}_F,
\end{align}
where the partial identity operator is defined as 
\begin{align}
 \hat{\mathbb{I}}_F=\sum_{m_F=-F}^F |F, m_F\rangle \langle F, m_F|.
\end{align}

In addition, we consider the interaction of the atoms in a ground state with a microwave field given by \cite{sinuco19}
\begin{align}
\hat{H}_{\mathrm{mw}}(t)&=\frac{\mu_B}{\hbar}\left(g_I \hat{\mathbf{I}}+g_J\hat{\mathbf{J}}\right)\cdot \mathbf{B}_{\mathrm{mw}}(t),
\end{align}
where $g_I$ and $g_J$ are the nuclear and electronic $g$-factors, respectively, with the corresponding angular momentum operators $\hat{\mathbf{I}}$ and $\hat{\mathbf{J}}$. In particular, for alkali atoms, for the ground state $^n S_{1/2}$ the orbital angular momentum is $L=0$, hence one can consider $\hat{\mathbf{J}}=\hat{\mathbf{L}}+\hat{\mathbf{S}}\approx \hat{\mathbf{S}}$. Moreover, $g_I\ll g_S$, which allows us to neglect the nuclear term. Thus, the microwave interaction Hamiltonian reduces to
\begin{align}
\hat{H}_{\mathrm{mw}}(t)&=\frac{\mu_Bg_S}{\hbar}\hat{\mathbf{S}}\cdot \mathbf{B}_{\mathrm{mw}}(t),
\end{align}
where the microwave field is described classically in the following form 
\begin{align}
\mathbf{B}_{\mathrm{mw}}(t)&=\tilde{\mathbf{B}}_{\mathrm{mw}}e^{-i\omega_{\mathrm{mw}}t}+\tilde{\mathbf{B}}_{\mathrm{mw}}^*e^{i\omega_{\mathrm{mw}}t},
\end{align}
with polarization $\tilde{\mathbf{B}}_{\mathrm{mw}}$ and frequency $\omega_{\mathrm{mw}}$.
The microwave field can be expressed in the spherical basis as
\begin{align}
\tilde{\mathbf{B}}_{\mathrm{mw}}&=\beta_{\mathrm{mw}}^0 \mathbf{e}_0 + \beta_{\mathrm{mw}}^+\mathbf{e}_+  +\beta_{\mathrm{mw}}^-\mathbf{e}_-,
\end{align}
where $\mathbf{e}_0=\mathbf{e}_z$ and $\mathbf{e}_\pm=\mp(\mathbf{e}_x+i\mathbf{e}_y)/\sqrt{2}$, and the magnetic field amplitudes are $\beta_{\mathrm{mw}}^0=\beta_{\mathrm{mw}~z}$ and 
 $\beta_{\mathrm{mw}}^\pm=\mp(\beta_{\mathrm{mw}~x}\pm i\beta_{\mathrm{mw}~y})$.
Therefore, the dynamics given by the Liouville equation is 
\begin{align}
\frac{d\hat{\rho} }{dt}=\frac{i}{\hbar}[\hat{\rho},\hat{H}_0+
\hat{H}_{\mathrm{mw}}(t)],\label{eq:Liouville}
\end{align}
in which its elements are $\rho_{Fm,F'm'}=\langle Fm|\hat{\rho}|F'm'\rangle$, where $F,F'\in\{1,2\}$.
Transforming into the rotating frame $\tilde{\rho}_{Fm,F'm'}=\rho_{Fm,F'm'}e^{i(F-F')\omega_{\mathrm{mw}}t}$, 
% for $i\neq j$,
we show that as in manifold two level optical system \cite{Steck15},
can be expressed in the rotating frame as
% \begin{align}
% \frac{d\tilde{\rho}_{\alpha m_\alpha, \beta m_\beta} }{dt}=&-\frac{i}{2}\sum_q\left[\delta_{\alpha,F'}\sum_{m_F} \Omega(m_\alpha,m_F)\tilde{\rho}_{Fm_F,\beta m_\beta}\right.\nonumber\\
% &-\delta_{F,\beta}\sum_{m'_F}\Omega(m'_F,m_\beta) \tilde{\rho}_{\alpha m_\alpha,F' m'_F}\nonumber\\
% &+\delta_{\alpha,F}\sum_{m'_F}\Omega^*(m'_F,m_\alpha) \tilde{\rho}_{F' m'_,\beta m_\beta}\nonumber\\
% &+\left.\delta_{F',\beta}\sum_{m_F}\Omega^*(m_\beta,m) \tilde{\rho}_{\alpha m_\alpha,F m}\right]\nonumber\\
% &+i(\delta_{\alpha,F'}\delta_{F,\beta}-\delta_{\alpha,F}\delta_{F',\beta})\Delta \tilde{\rho}_{\alpha m_\alpha, \beta m_\beta}
% \end{align}
\begin{align}
\frac{d\tilde{\rho}_{\alpha m_\alpha,\beta m_\beta} }{dt}&=i(\alpha-\beta)\Delta\tilde{\rho}_{\alpha m_\alpha,\beta m_\beta}\nonumber\\
&+i\delta_{F'\beta}\sum_{m,q}\Omega_{Fm,F'm_\beta}^q\tilde{\rho}_{\alpha m_\alpha F m} \nonumber\\
&-i\delta_{\alpha F}\sum_{m',q}\Omega_{Fm_\alpha,F'm'}^q\tilde{\rho}_{F' m' \beta m_\beta} \nonumber\\
&+i\delta_{F\beta}\sum_{m',q}\Omega_{F'm',Fm_\beta}^q\tilde{\rho}_{\alpha m_\alpha F' m'} \nonumber\\
&-i\delta_{\alpha F'}\sum_{m,q}\Omega_{F'm_\alpha,Fm}^q\tilde{\rho}_{F m \beta m_\beta},
\end{align}
where we have defined the microwave detuning as $\Delta=E_2-E_1 - \omega_{\mathrm{mw}}$ and the Rabi frequencies as
\begin{align}
\Omega_{Fm,F'm'}^q=\frac{\mu_B g_S}{\hbar}\frac{\mu_{Fma,F'mb}^q}{\hbar}\ \beta_{\mathrm{mw}}^q,
\end{align}
with $\mu_{Fm,F'mF'}^q=\langle F, m_F|\hat{S}^q|F', m_F'\rangle$.
The dynamics can be written in terms of an effective Hamiltonian as
\begin{align}
\frac{d\tilde{\rho} }{dt}=\frac{i}{\hbar}[\tilde{\rho},\hat{H}_{\mathrm{mw}}^{\mathrm{eff}}],
\end{align}
where we have defined $\hat{H}_{\mathrm{mw}}^{\mathrm{eff}}=\hat{H}_{0}^{\mathrm{eff}}+\hat{H}_{S}$ with
\begin{align}
\hat{H}_{0}^{\mathrm{eff}}=-\begin{bmatrix}
    0 & 0 & 0 &0 & 0 & 0 &0 & 0  \\
    0 & 0 & 0 &0 & 0 & 0 &0 & 0 \\
    0 & 0 & 0 &0 & 0 & 0 &0 & 0 \\
    0 & 0 & 0 &\Delta&0 & 0 & 0 &0\\
    0 & 0 & 0 &0&\Delta & 0 & 0 &0\\
    0 & 0 & 0 &0&0 & \Delta & 0 &0\\
    0 & 0 & 0 &0&0 & 0 & \Delta &0\\
    0 & 0 & 0 &0&0 & 0 & 0 &\Delta\\
\end{bmatrix},
\end{align}
whereas the microwave field Hamiltonian can be expressed as
\begin{align}
\hat{H}_{S}&=\Omega_\pi \hat{S}_z +\Omega_{\sigma+} \hat{S}_{\sigma+}+\Omega_{\sigma-} \hat{S}_{\sigma-},
\end{align}
% where the new spin operators are 
% defined in section \ref{app:MWRotFrame}.
where longitudinal spin operator is
\begin{align}
\hat{S}_{z}=\begin{bmatrix}
    0 & 0 & 0 &0 & \frac{\sqrt{3}}{4} & 0 &0 & 0  \\
    0 & 0 & 0 &0 & 0 & \frac{1}{2} &0 & 0 \\
    0 & 0 & 0 &0 & 0 & 0 &\frac{\sqrt{3}}{4} & 0 \\
    0 & 0 & 0 &0&0 & 0 & 0 &0\\
    \frac{\sqrt{3}}{4} & 0 & 0 &0&0 & 0 & 0 &0\\
    0 &  \frac{1}{2} & 0 &0&0 & 0 & 0 &0\\
    0 & 0 & \frac{\sqrt{3}}{4} &0&0 & 0 & 0 &0\\
    0 & 0 & 0 &0&0 & 0 & 0 &0\\
\end{bmatrix},\label{eq:Sz}
\end{align}
the circular positive operator is 
\begin{align}
\hat{S}_{\sigma_+}=\begin{bmatrix}
    0 & 0 & 0 & \frac{\sqrt{3}}{2} & 0 &0 &0& 0  \\
    0 & 0 & 0 & 0 & \frac{\sqrt{6}}{4} &0 &0& 0 \\
    0 & 0 & 0 & 0 & 0 & \frac{\sqrt{2}}{4} &0&0 \\
    0 & 0 & 0 & 0 & 0 &0\\
    0 & 0 & 0 &0&0 & 0 & 0 &0\\
    -\frac{\sqrt{2}}{4} & 0 & 0 &0&0 & 0 & 0 &0\\
    0 & - \frac{\sqrt{6}}{4} & 0 &0&0 & 0 & 0 &0\\
    0 & 0 & -\frac{\sqrt{3}}{2} &0&0 & 0 & 0 &0\\
\end{bmatrix},\label{eq:Sp}
\end{align}
and the negative circular operator is
\begin{align}
\hat{S}_{\sigma_-}=\begin{bmatrix}
    0 & 0 & 0 &0 & 0&-\frac{\sqrt{2}}{4} & 0 &0   \\
    0 & 0 & 0 &0 & 0& 0 & -\frac{\sqrt{6}}{4} &0 \\
    0 & 0 & 0 &0 & 0& 0 & 0 &-\frac{\sqrt{3}}{2} \\
    \frac{\sqrt{3}}{2} & 0 & 0 &0&0 & 0 & 0 &0\\
    0 &  \frac{\sqrt{6}}{4} & 0 &0&0 & 0 & 0 &0\\
    0 & 0 & \frac{\sqrt{2}}{4} &0&0 & 0 & 0 &0\\
    0 & 0 & 0 &0&0 & 0 & 0 &0\\
    0 & 0 & 0 &0&0 & 0 & 0 &0\\
\end{bmatrix}.\label{eq:Sn}
\end{align}

The dynamics describe the Rabi oscillations induced by the microwave field coupling the two hyperfine ground states of alkali atoms.

\subsection{Spin dynamics with the radio-frequency fields\label{Sup:rfDynamics}}
Now we add the interaction of atoms with a radio-frequency dressing field and external static fields 
% such that the hamiltonian interaction is 
\begin{equation}
\mathbf{B}=(B_{\mathrm{rf}} \cos{\omega_{\mathrm{rf}} t} +B_x^{\mathrm{ext}})\mathbf{e}_x+B_y^{\mathrm{ext}}\mathbf{e}_y+(B_{\mathrm{dc}}+B_z^{\mathrm{ext}})\mathbf{e}_z,\label{eq:Blab}
\end{equation}
where $B_{\mathrm{rf}}$ and $\omega_r$ are the amplitude and frequency of the rf driving field and $B_{\mathrm{dc}}$ is the static field along the longitudinal direction which are experimentally controlled. Additionally, we have the external fields
$B_i^{\mathrm{ext}}$ with $i=x,y$ and $z$, which originate from external sources. Therefore, 
the magnetic field interaction  is given by $\hat{H}_B=(\mu_Bg_F/\hbar)\hat{\mathbf{F}}\cdot\mathbf{B}$, where $\mu_B$ and $g_F$ correspond to the Bohr magnetron and $g_F$-factor, respectively, such that 
\begin{align}
\hat{H}_B(t)&=(\Omega_{\mathrm{rf}}\cos(\omega_{\mathrm{rf}} t)+\Omega_{x}^{\mathrm{ext}})\hat{F}_x
 + \Omega_y^{\mathrm{ext}}\hat{F}_y\nonumber\\&+(\Omega_{\mathrm{dc}}+\Omega_z^{\mathrm{ext}})\hat{F}_z,\label{eq:HB_int}
\end{align}
with $g'_F=g_F/\hbar$ and $\Omega_i=\mu_B g'_F B_i$ with $i=\mathrm{rf},\mathrm{dc},x,y$ and $z$.

%  We are interested in the situation where microwave frequency is resonant to the hiperfine splitting  and the radio frequency dressing field correspond to a slow varying field compared to the microwave field i.e. $\omega_{\mathrm{MW}}\gg \omega_{rf}$. Therefore, one can consider that the total interaction hamiltoninan is
Therefore,  the total interaction of the atoms with the microwave and the external magnetic fields dressed by the rf is described by the Hamiltonian
\begin{align}
\hat{H}_{T}(t)&=\hat{H}_{\mathrm{mw}}^{\mathrm{eff}}+\hat{H}_{B}(t).
\end{align}

% Thus, the dynamics is now given by the Liouville equation
% \begin{align}
% \frac{d\tilde{\rho} }{dt}=\frac{i}{\hbar}[\tilde{\rho},\hat{H}_{T}(t)]
% \end{align}
In addition to this term that contributes to the coherent dynamics, we need to include the relaxation term which models atom-atom collisions and atom-wall collisions, as well as the pumping rate which describes the state preparation by an pump beam. To do so, we introduce the standard input-output dynamics such that
\begin{align}
\frac{d\tilde{\rho} }{dt}=\frac{i}{\hbar}[\tilde{\rho},\hat{H}_{T}(t)]-\Gamma_p(t)(\tilde{\rho}-\rho_\mathrm{in}) - \gamma (\tilde{\rho}-\rho_{0}),
\end{align}
where $\Gamma_p(t)$ correspond to the pumping rate, which in general can be time dependent (for instance the Amplitude modulation employed to do the state preparation), and $\gamma$ represents the relaxation rate due to collisions.

For the 8 levels of the two hyperfine ground states,
 one can describe the dynamics above
 in the Liouville space, as in the case of ref.\cite{Floquet21}, by defining $\mathbf{X}=(\tilde{\rho}_{11},\tilde{\rho}_{12},\cdots,\tilde{\rho}_{88})$ with dimension $d_{\rho}=64$, such that 
% define indexing
\begin{align}
\frac{d\mathbf{X}(t)}{dt}=&(\mathbf{M}(t)-\Gamma_p(t)-\gamma)\ \mathbf{X}(t)+ \Gamma_{p} (t)\mathbf{X}_\mathrm{in}+\gamma\mathbb{X}_{0},\label{eq:Liouville_dyn}
\end{align}
where
\begin{align}
\mathbf{M}(t)=&\mathcal{L}[\hat{H}_{\mathrm{mw}}^{\mathrm{eff}}+\hat{H}_{B}(t)] +\mathcal{R}[\hat{H}_{\mathrm{mw}}^{\mathrm{eff}}+\hat{H}_{B}(t)],
\end{align}
in which $\mathcal{L}(\hat{O})$ and $\mathcal{R}(\hat{O})$ represents the action of the operator $\hat{O}$ to the left and to the right, respectively.

The dynamics of the radio-frequency dressed states are dominated by the 
time dependence of driving rf field $\mathbf{M}(t)$ and the amplitude modulation of the synchronous pumping determined by $\Gamma_p(t)$. On one hand, since the magnetic field in eq.~(\ref{eq:Blab}) can be harmonically decomposed as $\mathbf{B}(t)=\mathbf{B}^{(0)}+\mathbf{B}^{(1)} e^{i\omega_{\mathrm{rf}} t} +\mathbf{B}^{(-1)}e^{-i\omega_{\mathrm{rf}} t}$, therefore the magnetic interaction in eq.~(\ref{eq:HB_int})  can be also written as
\begin{align}
\mathbf{M}(t)=&\mathbf{M}^{(0)}+\mathbf{M}^{(1)}e^{i\omega_{\mathrm{rf}} t}+\mathbf{M}^{(-1)}e^{-i\omega_{\mathrm{rf}} t}.
\end{align}

Similarly, when the pump amplitude is modulated; for instance as a square-wave signal, the the pumping rate is decomposed as 
\begin{align}
\Gamma_p(t)=\Gamma_p^{(0)}+& \Gamma_p^{(1)} e^{i\omega_{\mathrm{rf}} t} +\Gamma_p^{(-1)}e^{-i\omega_{\mathrm{rf}} t} \nonumber\\
+& \Gamma_p^{(2)} e^{2i\omega_{\mathrm{rf}} t} +\Gamma_p^{(-2)}e^{-2i\omega_{\mathrm{rf}} t}+\cdots,\label{eq:Ypdecomp}
\end{align}
such that
\begin{align}
\Gamma_p^{(0)}&=\Gamma_b\ d,\ 
\Gamma_p^{(n)}=\Gamma^{(-n)}=\frac{\Gamma_b}{n\pi}\sin(n\pi d),
\end{align}
where $d$ corresponds to the duty cycle of the carrier wave.
The general description in eq.~(\ref{eq:Ypdecomp}) can simulate a broad range of time dependent pumping rates with different spectral decomposition e.g. sine, sawtooth etc. This harmonic decomposition of the dynamics generator, leads to a Floquet expansion to obtain a stationary solutions of its harmonics \cite{Floquet21}. In next section we present a brief description of the Floquet expansion in order to find the solution of the atomic dynamics.

\subsection{Floquet expansion\label{Sup:FloquetExpan} }
Given the harmonic feature of the dynamics generator, one can assume that the solution of the atomic state can be expressed in terms of the Floquet expansion $\mathbf{X}(t)=\sum_n\mathbf{X}^{(n)}(t) e^{in\omega_{\mathrm{rf}} t}$ so that one has to determine the amplitude of the harmonics $\mathbf{X}^{(n)}(t)$ to obtain the complete solution. Substituting this expansion into eq.~(\ref{eq:Liouville_dyn}), yields the recursive formula that can be written in the matrix form as
\begin{align}
\frac{d\mathbb{X}_F(t)}{dt}=&[\tilde{\mathbb{C}}-\mathbbm{\Gamma}]\ \mathbb{X}_F+ \mathbbm{\Gamma}_\mathrm{in} \mathbb{X}_\mathrm{in}+\mathbbm{\Lambda}_\mathrm{rel} \mathbb{X}_{0},
\end{align}
where $\mathbb{X}_F=(\mathbf{X}^{(-Q)},\allowbreak\cdots,\allowbreak\mathbf{X}^{(-n)},\allowbreak\cdots,\allowbreak\mathbf{X}^{(-1)},\allowbreak\mathbf{X}^{(0)},\allowbreak\mathbf{X}^{(1)},\allowbreak\cdots,\mathbf{X}^{(n)},\allowbreak\cdots,\allowbreak\mathbf{X}^{(Q)})^T$ correspond to the vector of harmonic amplitudes of the spin, with cutting frequency $Q$ and the dynamics generators are
\begin{align}
\tilde{\mathbb{C}}_{nm}=\begin{cases}
      \mathbf{C}^{(0)}-in\omega_\mathrm{rf} \mathbf{I}, & \text{for}\ n=m, \\
      \mathbf{C}^{(\pm1)}, & \text{for}\ m=n\mp1 \\
      0, & \text{otherwise}, \\
    \end{cases}
\end{align}
whereas the pump matrix elements $(\mathbbm{\Gamma})_{nm}=\Gamma_p^{(n-m)}\mathbf{I}_{3\times3}$
and 
\begin{align}
% (\mathbbm{\Gamma})_{nm}&=\Gamma_p^{(n-m)}\mathbf{I}_{3\times3}\\
(\mathbbm{\Gamma}_\mathrm{in})_{nm}&=\begin{cases}
      \Gamma_p^{(n)}\mathbf{I}_{3\times3}, & \text{for}\ n=m, \\
      0, & \text{otherwise}, \\
    \end{cases}
\end{align}
with the input vector $(\mathbb{P}_\mathrm{in})_n=\mathbf{P}^\mathrm{in}$.

\begin{figure}[t!]
\begin{overpic}[width=8.6cm]{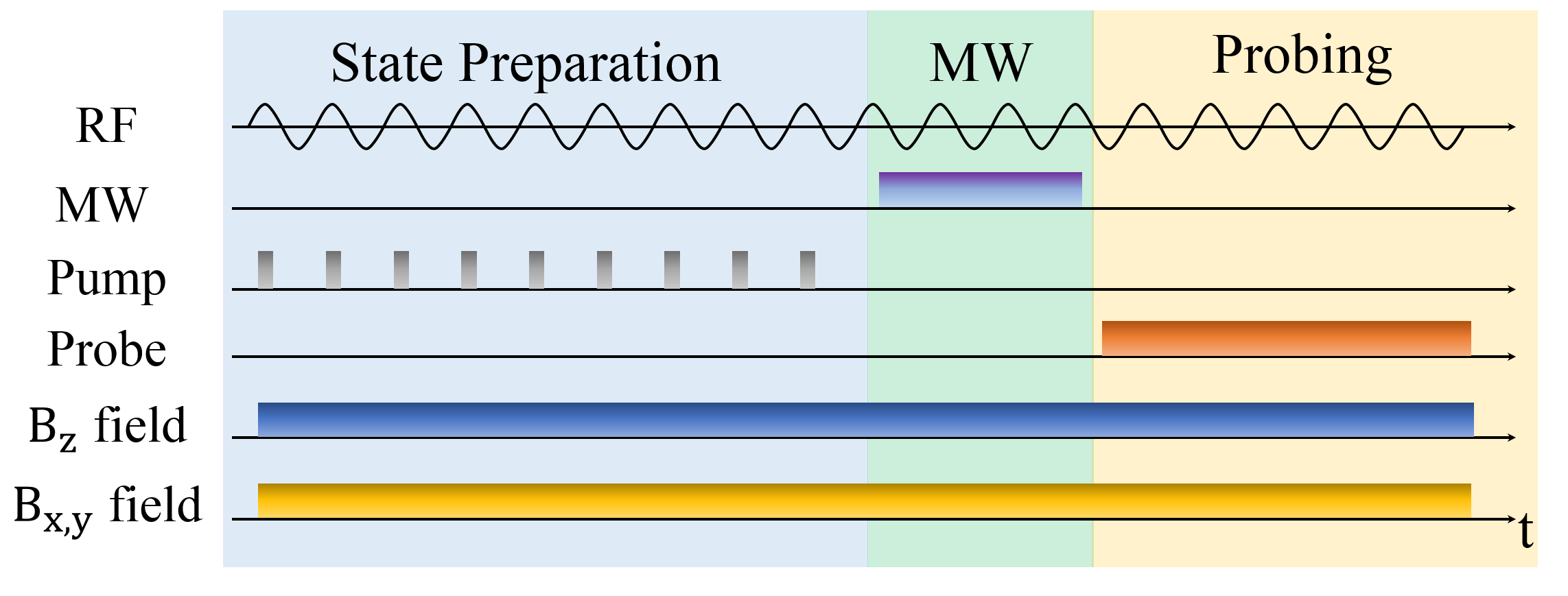} \end{overpic}
\caption{Sequence to produce the mw spectroscopy. The sequence is composed of three separate stages. First the process is initiated by the state preparation process where the rf-dressed atoms are synchronously pumped to prepare an aligned state. This is then followed by switching all of the optical fields off and applying a short mw pulsed tuned to the ground hyperfine splitting to induce a population transfer. The atomic dynamics are then probed by an off-resonant probe beam to measure the Voigt effect.}
\label{fig:mw_sequence}
\end{figure}

The magnetometer microwave spectrum has three stages: synchronous pumping, microwave interaction and probing, which are shown schematically in Fig.~\ref{fig:mw_sequence}. Each sequence is characterised by the following equation of motion
\begin{align}
&\frac{d\mathbb{X}_F(t)}{dt}\\
&=\begin{cases}
      [\tilde{\mathbb{C}}_0-\mathbbm{\Gamma}_T]\ \mathbb{X}_F+ \mathbbm{\Gamma}_\mathrm{in} \mathbb{X}_\mathrm{in}+\mathbbm{\Lambda}_\mathrm{rel} \mathbb{X}_{0}, & \text{Pump Cycle,} \\[5pt]
     [\tilde{\mathbb{C}}-\mathbbm{\Lambda}_\mathrm{rel}]\ \mathbb{X}_F+\mathbbm{\Lambda}_\mathrm{rel} \mathbb{X}_{0}, & \text{MW Cycle,} \\[5pt]
     [\tilde{\mathbb{C}}_0-\mathbbm{\Lambda}_\mathrm{rel}]\ \mathbb{X}_F+\mathbbm{\Lambda}_\mathrm{rel} \mathbb{X}_{0}, & \text{Probe Cycle,}
    \end{cases}\nonumber
\end{align}
where we have defined $\tilde{\mathbb{C}}_0=\tilde{\mathbb{C}}(H_\mathrm{mw}=0)$.
During the pumping stage, as before, a steady state is reached. This is given by
\begin{align}
 \mathbb{X}^{\mathrm{Pump}}_F=&-[\tilde{\mathbb{C}}_0-\mathbbm{\Gamma}_T]^{-1}(\mathbbm{\Gamma}_\mathrm{in} \mathbb{X}_\mathrm{in}+\mathbbm{\Lambda}_\mathrm{rel}\mathbb{X}_{0}).\label{eq:pump_mw}
\end{align}
After reaching the steady state, the pump pulse is switched off, which is immediately followed by an application of a microwave pule. During microwave cycle, %the system evolves freely which is described by the following equation of motion 
% \begin{align}
% \left.\frac{d\mathbb{X}_F(t)}{dt}\right|_{\mathrm{MW}}=[\tilde{\mathbb{C}}-\mathbbm{\Lambda}_\mathrm{rel}]\ \mathbb{X}_F+\mathbbm{\Lambda}_\mathrm{rel} \mathbb{X}_{0}.
% \end{align}
% Integrating the equation above allows us to compute the evolution of the density matrix after time, $t$, namely
the time evolution is

\begin{align}
\mathbb{X}^{\mathrm{mw}}_F(t)&=e^{(\tilde{\mathbb{C}}_{\mathrm{mw}}-\mathbbm{\Lambda}_\mathrm{rel})t}\mathbb{X}^{\mathrm{Pump}}_F\\
&+(\tilde{\mathbb{C}}_{\mathrm{mw}}-\mathbbm{\Lambda}_\mathrm{rel})^{-1}(e^{(\tilde{\mathbb{C}}_{\mathrm{mw}}-\mathbbm{\Lambda}_\mathrm{rel})t}-\mathbb{I})\mathbbm{\Lambda}_\mathrm{rel} \mathbb{X}_{0}.\nonumber
\end{align}
If the microwave is continuously interacting for an integer number of rf cycles $t_{\mathrm{mw}}=n/\omega_{\mathrm{rf}}$, then the state at a later time is simply $\mathbb{X}^{\mathrm{mw}}_F(t_{\mathrm{mw}})$.
% \begin{align}
% \mathbb{X}^{\mathrm{MW}}_F(t_{\mathrm{MW}})=&e^{(\tilde{\mathbb{C}}_{\mathrm{MW}}-\mathbbm{\Lambda}_\mathrm{rel})t_{\mathrm{MW}}}\mathbb{X}^{\mathrm{Pump}}_F\nonumber\\
% &+(\tilde{\mathbb{C}}_{\mathrm{MW}}-\mathbbm{\Lambda}_\mathrm{rel})^{-1}(e^{(\tilde{\mathbb{C}}_{\mathrm{MW}}-\mathbbm{\Lambda}_\mathrm{rel})t_{\mathrm{MW}}}-\mathbb{I})\mathbbm{\Lambda}_\mathrm{rel} \mathbb{X}_{0}.
% \end{align}
After the microwave pulse, a probe pulse is applied for a given time duration $t_{\mathrm{Probe}}$. During the probing, %the equation of motion reads
the solution is 
% \begin{align}
% \left.\frac{d\mathbb{X}_F(t)}{dt}\right|_{\mathrm{Probe}}=[\tilde{\mathbb{C}}-\mathbbm{\Lambda}_\mathrm{rel}]\ \mathbb{X}_F+\mathbbm{\Lambda}_\mathrm{rel} \mathbb{X}_{0},
% \end{align}
% which has a solution
\begin{align}
\mathbb{X}^{\mathrm{Probe}}_F(t)&=e^{(\tilde{\mathbb{C}}-\mathbbm{\Lambda}_\mathrm{rel})t}\mathbb{X}^{\mathrm{mw}}_F(t_{\mathrm{mw}})\nonumber\\
&+(\tilde{\mathbb{C}}-\mathbbm{\Lambda}_\mathrm{rel})^{-1}(e^{(\tilde{\mathbb{C}}-\mathbbm{\Lambda}_\mathrm{rel})t}-\mathbb{I})\mathbbm{\Lambda}_\mathrm{rel} \mathbb{X}_{0}.\label{eq:solFloquet}
\end{align}
The equation above is the solution from which we can determine the slow varying envelopes for the Voigt rotation
\begin{align}
\Braket{\hat{S}_z'(t)}&=\Braket{\hat{S}_z(t)}+
G_{F}^{(2)} S_y n_F \Braket{\hat{F}_x^2(t)-\hat{F}_y^2(t)},\label{eq:Voigt}
\end{align}
with its dynamics described by the second harmonic in the limit when no transverse fields are applied~\cite{Tadas19}. According to eq.~(\ref{eq:Voigt}), when the probe has no initial ellipticity, 
the Voigt rotation for  the ground state $F=1$ and $F=2$ are
% $ \Braket{\hat{F}_x^2(t)-\hat{F}_y^2(t)}=$
% where we compute the mode amplitude $h_2=(\mathbf{X}_{R}^{(2)})$
\begin{eqnarray}
\Braket{\hat{S}_z'(t)}_{F=1}&=
G_{F}^{(2)} S_y n_F[&\rho_{13}(t)+\rho_{31}(t)]\label{eq:VoigtF1},\\
\Braket{\hat{S}_z'(t)}_{F=2}&=
G_{F}^{(2)} S_y n_F\Big[&\sqrt{6} [\rho_{46}(t)+\rho_{64}(t)]\nonumber\\
& &+\sqrt{6} [\rho_{68}(t)+\rho_{86}(t)]\nonumber\\
& &+3 [\rho_{75}(t)+\rho_{57}(t)]\Big]\label{eq:VoigtF2},
\end{eqnarray}
where each term has a spectral expansion of the form $\rho_{ij}(t)=\rho_{ij}^{(0)}(t)+\rho_{ij}^{(1)}(t)e^{i\omega_\mathrm{rf}t}+\rho_{ij}^{(2)}(t)e^{2i\omega_\mathrm{rf}t}+\cdots+c.c.$.
Thus, by extracting the mode amplitude of the second harmonic in eq.~(\ref{eq:solFloquet})
% Since $(\mathbb{X}^{\mathrm{Probe}}_F(t))_m$
\begin{align}
\mathbf{X}_{R}^{(2)}=\frac{1}{2T}\int_0^T dt' \left[(\mathbb{X}^{\mathrm{Probe}}_F(t'))_{2}+(\mathbb{X}^{\mathrm{Probe}}_F(t'))_{-2}\right],\label{eq:XR2}
\end{align}
one can express the ellipticities in eq.~(\ref{eq:VoigtF2})
in terms of the solution in eq.~(\ref{eq:XR2}), since the second harmonic matrix elements are $\rho^{(2)}_{ij}(t)=[\mathbf{X}_{R}^{(2)}]_{d_{\rho}\times i +j}$.

\section{Experimental setup \label{Sup:exp_setup}}
Here we present some more details of our magnetically unshielded experimental setup is depicted in Fig.~\ref{fig:setup}~(a). 
% It follows closely the shielded version of the experimental setup described in ref.~\cite{Tadas19}. 
A paraffin coated $^{87}$Rb enriched vapour cell {\color{black}of diameter} $d=26$~mm and {\color{black}length} $l=75$~mm at room temperature, with a density of approximately $10^{10}$ atoms per cubic cm. The atomic state is dressed with a radio-frequency field which is generated by cosine-theta coil along x-axis. The atoms are dressed with a $\omega_\mathrm{rf}=2\pi\times~90$~kHz rf field and and coupled to a static field in the longitudinal direction. Three Helmholtz coils are used to actively compensate and stabilise the external magnetic field using an analog three channel PID controller driving a home made bipolar current source. The in-loop field is sensed by a three axis fluxgate magnetometer (Stefan-Mayer FLC3-70). The atomic state is prepared by a combination of co-propagating linearly polarized pump laser beam tuned to the $F=2$ to $F'=1$ transition of the D1 line and a linearly polarized repump laser tuned to $F=1$ to $F'=2$ of the D2 line. The atoms are stroboscopically pumped by modulating the pump amplitude with 10\% duty cycle in phase with the rf field with the repump set in cw mode.

% \begin{figure}[t!]
% \begin{overpic}[width=0.5\textwidth]{mw_setup.png}
% \end{overpic}
% \caption{Sketch of the experimental setup. Here a single mode polarisation maintaining fibre containing pump and repump beams pumps the dressed atoms to prepare them in an aligned state. A dipole antenna in close proximity to the vapour cell produces a short mw-pulse to induce a population transfer between the hyperfine ground states. Finally a linearly polarised probe beam interacts with the atoms through the Voigt effect which is analysed using a balanced homodyne detector with an IQ demodulator. }
% \label{fig:setup}
% \end{figure} 

For a complete characterization of the atomic state, we employ the Voigt rotation  as non-destructive measurement to probe the atoms in hyperfine ground state $F=1$ and $F=2$. To do so, the atoms are probed by a $45^\circ$ polarized laser beam relative to the pump polarization and tuned $-500$~MHz with respect to the $F=2$ to $F'=1$ transition and $F=1$ to $F'=1$ transition of the D1 line. After the interaction with the atoms, the light passes through a quarter waveplate and a polarizing beam splitter, which allows us to measure the Voigt rotation of the light polarization i.e.\ the ellipticity induced by the atoms. The light is detected on a balanced photodetector  where the detected signal $u(t)=g_{el} S_z(t)$ is proportional to the ellipticity in eq.~(\ref{eq:Voigt}) and the electronic gain $g_{el}$. As it was shown in ref.~\cite{Tadas19}, the ellipticity for the Voigt rotation produces a signal at the first and second harmonic of the radio-frequency dressing field such that $u(t)=m_0+m_1e^{i\omega_\mathrm{rf}t}+m_2e^{2i \omega_\mathrm{rf}t}+c.c$. This output signal is demodulated at the second harmonic using a home built IQ demodulator, from which we can extract its mode amplitude $m_2$. 

The microwave spectroscopy is performed when the magnetometer is tuned on the resonance, $B_z=\hbar\omega_\mathrm{rf}/\mu_Bg_F$, of the second harmonic of the Voigt rotation, cancelling the presence of any transverse field i.e.\ zeroing the first harmonic~\cite{Tadas19}.
%Description of the antena
The microwave field is generated using an rf generator (SRS SG380) with a frequency doubler (Minicircuits ZX90-2-36-S+) and an additional amplifier. The signal is coupled into a half-wave dipole antenna $L=\lambda/2=21.9$~mm where $L$ is the conductor length and $\lambda$ corresponds to $\Ket{F=1}\rightarrow\Ket{F=2}$ hyperfine microwave transition. The direction of the dipole antenna is transverse to the light propagation. 

\begin{figure}[t!]
\begin{overpic}[width=0.5\textwidth]{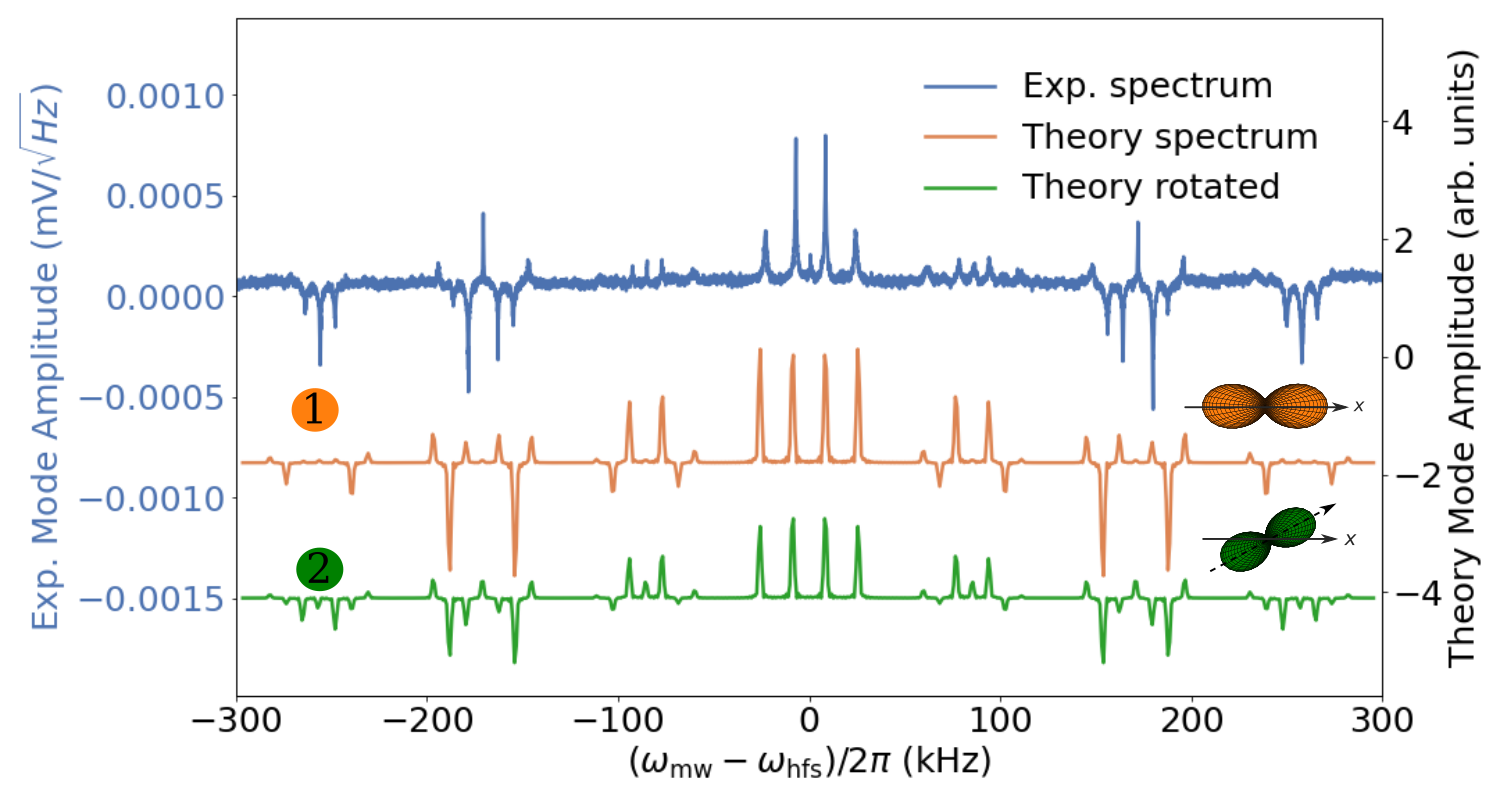}
\put(0,50){(a)}
\end{overpic}
\begin{overpic}[width=0.5\textwidth]{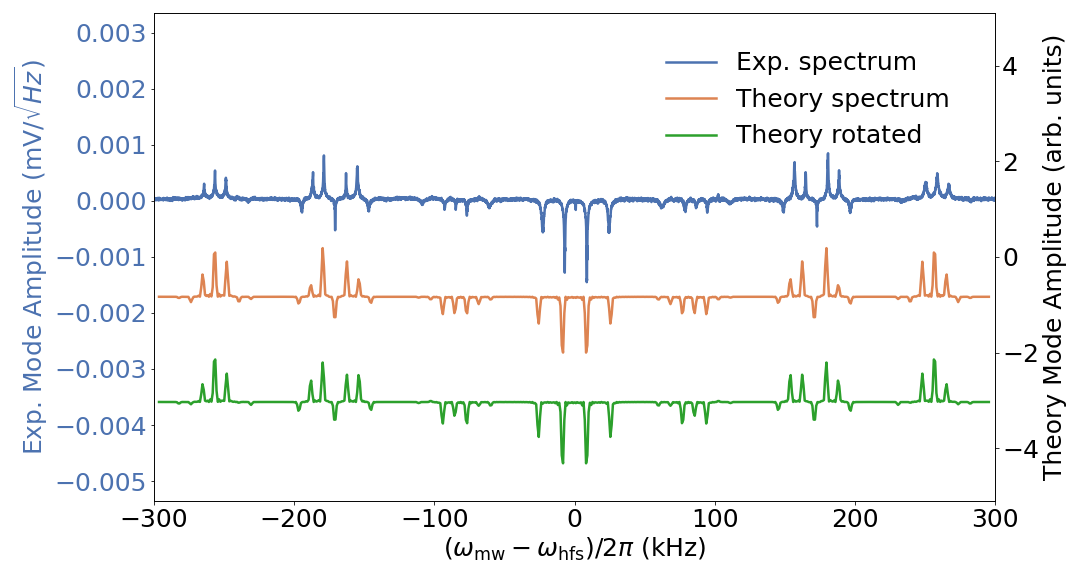}
\put(0,50){(b)}
\end{overpic}
\caption{Experimental and theoretical dressed microwave spectra of the $\mathrm{Re}(m_2)$ mode amplitude probing the $F=1$ manyfold following the sequence in Fig.~\ref{fig:cone_cw_mw}. (a) Repump ON. (b) Repump OFF. In both cases we considered an effective rotation of $35^\circ$.}
\label{fig:CW_MW_spectrum_F1}
\end{figure}

\section{Micro-wave spectroscopy probing $F=1$ state \label{supl:F1_spectrum}}
Complementing the observations of the mw spectroscopy  when we probed the state $F=2$, in this section we present the mw spectroscopy when the $F=1$ manifold is probed.
Figure~\ref{fig:CW_MW_spectrum_F1}(a) shows the mw spectrum when the repump is ON during the pumping process. Figure (b) shows the case when the repump is OFF such that the birefringent signal is acquires opposite polarity, since the residual population in $F=1$ can now be excited to $F=2$ ground state. It is worth noting that the birefringent signal mean value is around zero since its population is empty and initially prepared in $F=2$.

\begin{figure}[t!]
\begin{overpic}[width=0.5\textwidth]{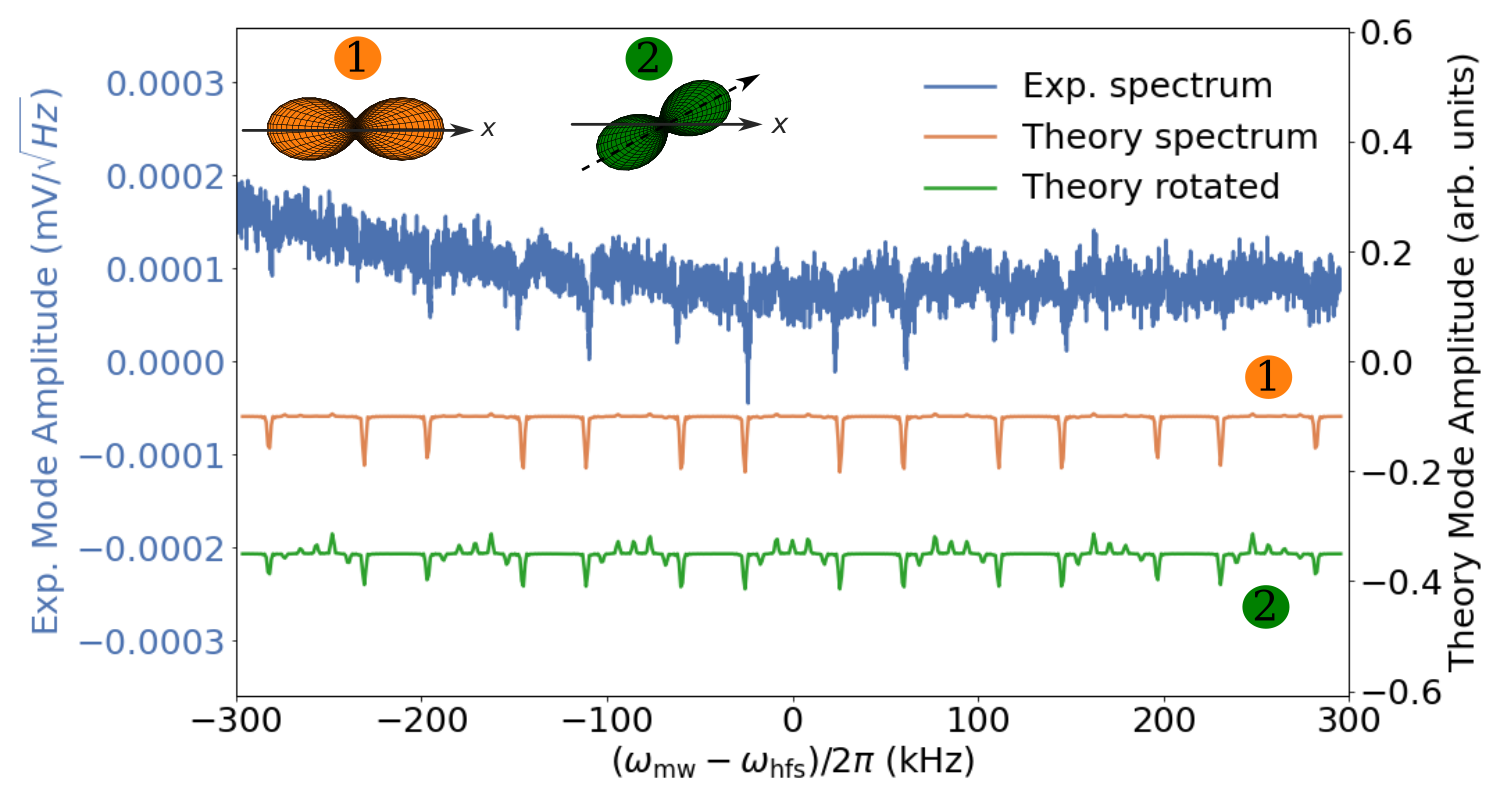}
\put(0,50){(a)}
\end{overpic}
\begin{overpic}[width=0.5\textwidth]{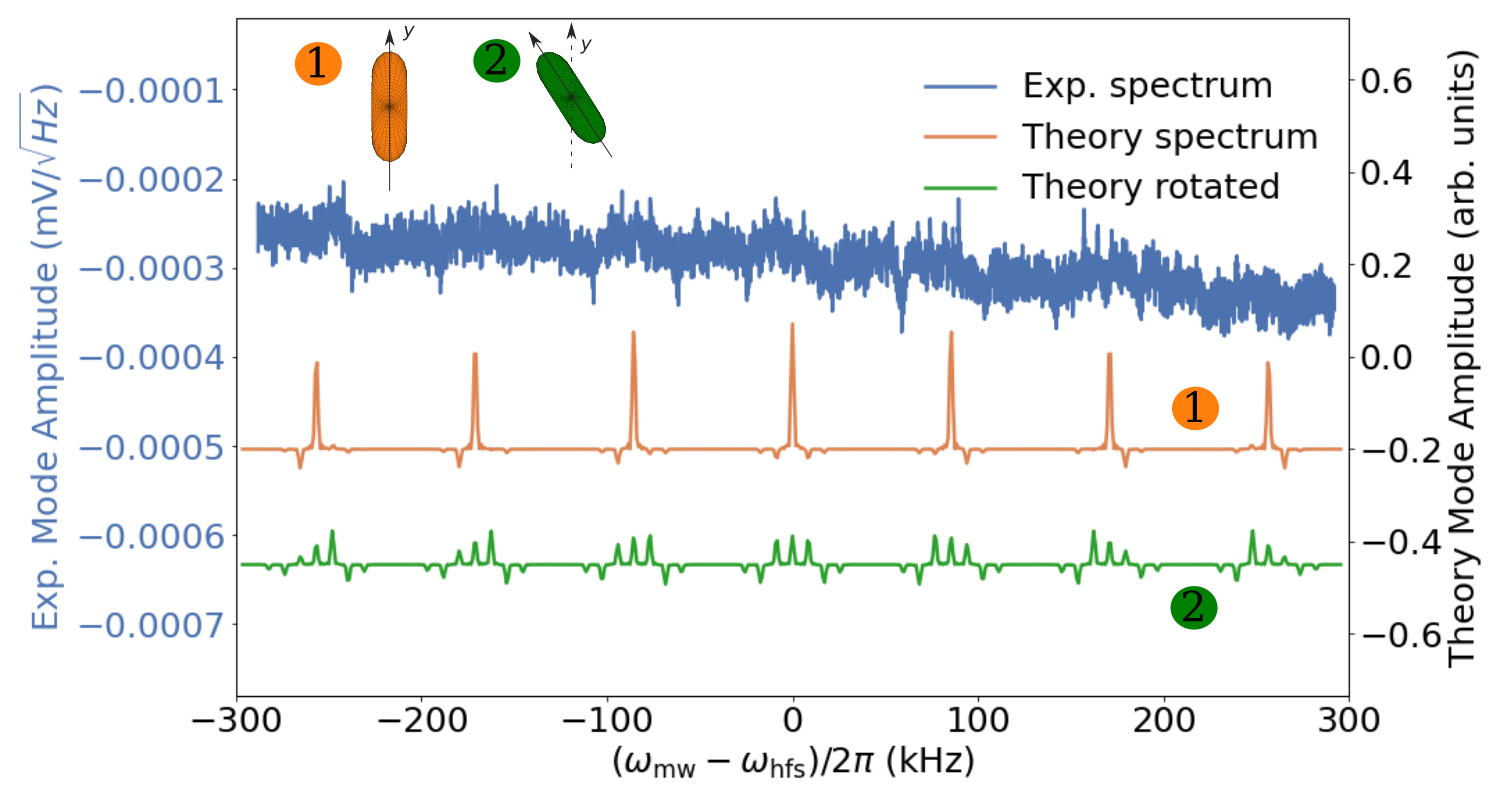}
\put(0,50){(b)}
\end{overpic}
\caption{Experimental and theoretical dressed microwave spectra of the $\mathrm{Re}(m_2)$ mode amplitude for sinchronously mw interaction. (a) Aligned state. (b) Clock state. In both cases we considered an effective rotation of $35^\circ$.}
\label{fig:PL_MW_spectrum_F1}
\end{figure}

Similarly to the case of probing $F=2$, in this case the propagation effects are also present in the pump-atom interaction
, degrading the efficiency the state preparation. Notice that in the case when the repump is ON, the extreme groups describes the populations observed experimentally, whereas for the case of no repump, the correction with the rotation does not appear to change the profile.

Now, implementing the synchronously mw spectroscopy, one can notice the appearance mainly of the extreme peaks of each group. However the contamination due to the propagation effects are noticeable, consistent with what is observed for $F=2$.

Although the mw spectra for $F=2$ have better signal-to-noise ratio, this observations show that the state characterisation can be done in both manifolds $F=1$ or $F=2$. Furthermore, the model allows to explore different scenarios, for instance, the mw spectroscopy of with arbitrary mw polarisation and pump amplitude modulations.

\end{document}